\documentclass[10pt]{iopart}
\usepackage{epsf,epsfig}
\usepackage{amsfonts}
\newcommand{\real}{{\rm I\!R}}
\newcommand{\be}{\begin{equation}}
\newcommand{\ee}{\end{equation}}
\newcommand{\bd}{\begin{displaymath}}
\newcommand{\ed}{\end{displaymath}}

\newcommand{\bnull}{\mbox{\boldmath ${0}$ }}

\newcommand{\bra}{\langle}
\newcommand{\ket}{\rangle}

\newcommand{\BN}{\vspace*{3mm}\begin{quote}\hrule\sl $~$\\ NOTE: }
\newcommand{\EN}{\\ \hrule\end{quote}\vspace*{3mm}}

\newcommand{\room}{\rule[-0.3cm]{0cm}{0.8cm}}

\newcommand{\vsp}{\vspace*{3mm}}
\newcommand{\bdm}{\begin{displaymath}}
\newcommand{\edm}{\end{displaymath}}
\newcommand{\bea}{\begin{eqnarray}}
\newcommand{\eea}{\end{eqnarray}}
\newcommand{\sgn}{~{\rm sgn}}

\newcommand{\bigbra}{\left\langle\room}
\newcommand{\bigket}{\right\rangle\room}

\newcommand{\order}{{\cal O}}
\newcommand{\minus}{\!-\!}
\newcommand{\plus}{\!+\!}

\newcommand{\bx}{\mbox{\boldmath $x$}}

\newcommand{\bK}{\mbox{\boldmath $K$}}
\newcommand{\bM}{\mbox{\boldmath $M$}}

\newcommand{\bL}{\mbox{\boldmath $L$}}

\newcommand{\bphi}{\mbox{\boldmath $\phi$}}
\newcommand{\bfeta}{\mbox{\boldmath $\eta$}}

\newcommand{\bmu}{\mbox{\boldmath $\mu$}}

\newcommand{\unity}{{\bf 1}\hspace{-1mm}{\bf I}}
\begin{document}
\title[Sequence selection and compactification in mean-field
hetero-polymers]{Coupled dynamics of sequence selection and
compactification in mean-field hetero-polymers}
\author{
H Chakravorty$^\dag$,  A C C Coolen$^\dag$ and D
Sherrington$^\ddag$}
\address{\dag Department of Mathematics, King's College London,
            The Strand, London WC2R 2LS, UK\\[1mm]
          \ddag  Department of Physics -- Theoretical Physics, University of Oxford, 1 Keble Road, Oxford OX1 3NP, UK}
\begin{abstract}
We study a simple solvable model describing the genesis of monomer
sequences for hetero-polymers (such as proteins), as the result of
the equilibration of a slow stochastic genetic selection process
which is assumed to be driven by the competing demands of
functionality and reproducibility of the polymer's folded
structure. Since reproducibility is defined in terms of properties
of the folding process, one is led to the analysis of the coupled
dynamics of (fast) polymer folding and (slow) genetic sequence
selection. For the present mean-field model this analysis can be
carried out using the finite-dimensional replica method, leading
to exact results for (first- and second-order) transitions and to
rich phase diagrams.
\end{abstract}


\section{Introduction}

The functionality of a protein depends largely on the shape of its
3D native state. Determining this shape from a protein's sequence
of amino-acids is called the 'protein folding problem'. From a
medical point of view it is desirable to be able to solve the
related 'inverse folding problem': given a 3D structure, find a
sequence of amino-acids which would have this as its native state.
Unfortunately, X-ray crystallography (the standard tool for
determining native states) is very time-consuming, and will not
deliver the native states of the many amino-acid sequences
collected through the Human Genome Project for years to come
\cite{PDB,PIR}. Furthermore, it does not reveal the mechanics of
the folding process. Molecular dynamics simulations are also
extremely slow; only the first 10-50 $ns$ of protein folding
processes could so far be simulated
\cite{MDreview,Dill,Pande,Wang}. In spite of ongoing hardware
improvements, it will take long before real progress is made. This
points to the need for solvable mathematical models, designed to
capture the essentials of proteins, to shed more light on the
nature of the folding process and the role of its parameters.

The main complications in solving  protein models are the chain
constraint and the presence of disorder, embodied in the
amino-acid sequences (see e.g. \cite{GOT,GOP}). Natural amino-acid
sequences have evolved genetically, driven mainly by the two
(competing) demands of structure reproducibility and
functionality. They are not random (random hetero-polymers usually
do not fold into unique shapes), whereas most disordered systems
techniques are based on exploiting self-averaging properties of
random disorder. It is not yet understood what distinguishes a
random amino-acid sequence from a natural one. Our objective  is
to define and study a solvable model describing the interplay
between the genetic forces and folding processes, as a first step
towards understanding the genesis and statistics of natural
amino-acid sequences.

Our model is a simple mean-field hetero-polymer whose microscopic
state is described by two degrees of freedom per monomer:
$\phi_i\in[0,2\pi]$, giving the orientation of monomer $i$
relative to the backbone, and $\eta_i\in\{-1,1\}$, giving its
polarity (i.e. hydrophobic vs. hydrophilic)\footnote{Thus, in
addition to the mean-field nature of the forces and the absence of
a chain constraint, our model simplifies biological reality
further by reducing the orientation degrees of freedom of
individual amino-acids to one, and the characterization of the
physical properties of individual amino-acids to a binary
number.}. The evolution of the orientation variables  represents
the folding; that of the polarity variables (which involves
changing monomer species) represents genetic evolution. The latter
clearly takes place over much larger time-scales than the former,
and is the result of an interplay between minimizing an energetic
cost determined by the average folding quality, constraints on the
monomer composition, and noise. Models with adiabatically
separated time-scales have been studied and solved using
finite-dimensional replica theories, but so far mainly in the
context of neural networks and spin-glasses
\cite{CPS,PCS,DFM,PS,Caticha,Jongenetal2,MourikCoolen,UezuCoolen}.
We show how these techniques can also be used to study
analytically the coupled dynamics of (fast) folding and (slow)
genetic sequence selection in our present model.

We first define our model and solve it in the stationary state,
describing equilibrium at the largest (genetic) time-scale,
 via the finite-dimensional replica method.
Analysis of the  resulting macroscopic
 equations leads to general results regarding the ground state
 and (first- and second-order) phase transitions.
 We inspect our equations more closely for specific (small) values
 of the number $q$ of allowed local monomer orientations, and generate
phase diagrams via a combination of analytical and numerical
techniques. These are found to be very rich, and involve various
types of single-state and multiple-state swollen and compact
phases.

\section{The model and its solution}

\subsection{Model definitions}

We study hetero-polymer models as in \cite{Skantzos}, consisting
of $N$ monomers labelled $i=1\ldots N$. The spatial degrees of
freedom of the monomers are angles $\phi_{i} \in C\subseteq
[0,2\pi]$, describing their orientations around a polymer
backbone; each monomer $i$ can take only a finite number of
positions on a circle. The physical properties of a monomer $i$
are defined by its species (i.e. amino-acid type). Here we
restrict ourselves by taking into consideration only a monomer's
polarity $\eta_{i}=\real$, where $\eta_{i}>0$ for hydrophylic
monomers and $\eta_{i}<0$ for hydrophobic ones. A monomer sequence
is thus fully described by the vector
$\bfeta=(\eta_{1},...,\eta_{N})\in\real^N$, whereas the spatial
configuration (or `conformation') of the system is described by
the vector $\bphi=(\phi_{1},...,\phi_{N})\in C^N$.

On the time-scales of a single evolutionary generation the
sequence $\bfeta$ is fixed, and the only allowed process is
folding, i.e. evolution of the $\bphi$. We consider here only the
dominant folding force: compactification of the polymer via
`shielding' from the solvent of its hydrophobic monomers. A simple
well-known type of phenomenological Hamiltonian to describe this
effect is (see e.g. \cite{Garel,SG,Stafos})
\be
H({\bphi},\bfeta)=
-\frac{J}{N}\sum_{ij}\eta_{i}\eta_{j}~\delta[\phi_{i}-\phi_j]
\label{eq:fastham} \ee  with $J>0$. The rationale is that
efficient shielding of hydrophobic monomers and exposure of
hydrophilic ones requires {\em separation} of the species; ideally
with the two polarity types oriented at opposite sides of the
chain (here a single bend of the chain would ensure shielding).
The Hamiltonian (\ref{eq:fastham}) punishes monomer pairs with
opposite polarity but identical orientation relative to the
backbone (which make polarity-based compactification more
difficult), and favors pairs with identical polarity and
orientation. The mathematical simplifications induced by the
separable character and the infinite range of the monomer
interactions in (\ref{eq:fastham}) will allow us to solve the
super-imposed slow genetic process to be introduced below.

For a fixed realization of the sequence $\bfeta$, the equilibrium
statistics of the orientation variables, at temperature
$T=\beta^{-1}$, are characterized by the partition function
\be
Z[\bfeta]={\rm Tr}_{\bphi}~e^{-\beta H(\bphi,\bfeta)}
\label{eq:Zphi} \ee
 In addition to the folding process we now
introduce a stochastic dynamics for the sequences $\bfeta$, which
reflects the demands of structure reproducibility and structure
functionality. The former is measured by the achieved degree of
species separation, i.e. by (\ref{eq:fastham}). Upon describing
the degree of functionality of a sequence $\bfeta$  by a potential
$V(\bfeta)$, we arrive at the Langevin equation
\be
\frac{d}{dt}\eta_i=-\frac{\partial}{\partial\eta_i} \left\{
H(\bphi,\bfeta)+V(\bfeta)\right\} + \xi_i \label{eq:slow_Langevin}
\ee with zero-average Gaussian random forces $\xi_i(t)$, obeying
$\bra\xi_i(t)\xi_j(t^\prime)\ket=2\tilde{T}\delta_{ij}\delta(t-t^\prime)$
(which introduces a second `temperature'
$\tilde{T}=\tilde{\beta}^{-1}$). Our further discussion will be
restricted to the stationary state of the genetic process
(\ref{eq:slow_Langevin}). Since the genetic process is
adiabatically slow compared to the folding process, we may replace
(\ref{eq:slow_Langevin}) by its average over the equilibrium
folding statistics $p(\bphi|\bfeta)\sim e^{-\beta
H(\bphi,\bfeta)}$, i.e. \bd \frac{\partial
H(\bphi,\bfeta)}{\partial \eta_i}~~\to~~ \bra\frac{\partial
H(\bphi,\bfeta)}{\partial
\eta_i}\ket_{\bphi}=\frac{\partial}{\partial \eta_i}\left\{
-\frac{1}{\beta}\log Z[\bfeta]\right\}
 \ed
 The equilibrium measure
of the genetic process is now also of the Boltzmann form, with a
`genetic' Hamiltonian $H(\bfeta)$ which is the sum of the free
energy of the folding process, given the sequence $\bfeta$, and
the functionality potential $V(\bfeta)$), and characterized by an
associated genetic free energy per monomer $f_N$ (with the
`folding partition function' defined in (\ref{eq:Zphi})):
\be
H(\bfeta)=V(\bfeta)-\frac{1}{\beta}\log Z[\bfeta] ~~~~~~~~~~~
f_N=-\frac{1}{\tilde{\beta} N}\log {\rm Tr}_{\bfeta}~
e^{-\tilde{\beta}~H(\bfeta)} \label{eq:genetic_H} \ee We now make
a specific choice for the set $C$ of allowed single-monomer
angles, viz. $C=\{(2k\plus 1)\pi/q,~k=0,\ldots,q\minus 1\}$ ($q$
possible orientations per monomer), and we choose the simplest
non-trivial type of functionality potential
$V(\bfeta)=\sum_{i=1}^N \mu_i\eta_i$. In addition we switch to the
Ising version of (\ref{eq:genetic_H}), i.e. $\eta_i\in\{-1,1\}$
for all $i$ (as opposed to $\eta_i\in\real$)\footnote{Note that
this implies that the genetic dynamics (\ref{eq:slow_Langevin})
will be replaced by either a Glauber-type dynamics, or by a
`soft-spin' Langevin equation (involving the familiar parametrized
double-well potential for each $\eta_i$) from which the binary
version of the model can be obtained as a limit.}. In combination,
studying the stationary state at the largest (genetic) time-scale
has now been reduced to the calculation of the free energy per
monomer $f_N$: \bd f_N=-\frac{1}{\tilde{\beta} N}\log
\sum_{\bfeta}\!
 \left[\sum_{\bphi}e^{-\beta H(\bphi,\bfeta)}
 \right]^{\frac{\tilde{\beta}}{\beta}}
\!\!\! e^{-\tilde{\beta}\bmu\cdot\bfeta} ~~~~~~~~~~
H({\bphi},\bfeta)=
-\frac{J}{N}\sum_{ij}\eta_{i}\eta_{j}~\delta_{\phi_{i},\phi_j} \ed
\vspace*{-5mm}
\be
 \label{eq:theproblem} \ee with
$\bfeta\in\{-1,1\}^N$ and with $\bmu=(\mu_1,\ldots,\mu_N)$.

The energetic forces in our model favor species-based monomer
separation (due to the fast process) and species selection
controlled by the disorder variables $\{\mu_i\}$ (due to the slow
process). The ease with which these two aims can be achieved
simultaneously depends on the statistics of the single-site
genetic forces $\{\mu_i\}$; the relative importance of the two
objectives is controlled by the ratio $\tilde{\beta}/\beta$ of
associated noise levels. Entropic forces act in the opposite
direction, favoring randomly distributed monomer orientations and
species. The equilibrium state of the system must therefore be
characterized by disorder-averaged order parameters which measure
 the overall joint distribution of orientation- and species
variables along the chain.

\subsection{Solution via the finite-$n$ replica method}

Expression (\ref{eq:theproblem}), which is of the typical form
found for systems with disparate time-scales, can be evaluated
using the replica method. The temperature ratio
$n=\tilde{\beta}/\beta$ is first regarded as integer, which allows
us to write $Z^n[\bfeta]$ as the partition function of an $n$-fold
replicated system, followed by analytic continuation to
non-integer $n$ later (if needed). In the special case
$n\rightarrow{0}$ the polarity numbers $\{\eta_i\}$ reduce to
quenched disorder variables, and we return to a simplified version
of the model in \cite{Skantzos}. For $n=1$ one recovers the
annealed case (although the time-scales remain completely
disparate). The limit $n\rightarrow\infty$ corresponds to dominant
coupling of the two processes: here the polarity dynamics is fully
deterministic. When referring to temperature in the remainder of
this paper we will mean $T$, with $\tilde{T}$ linked to $T$ via
$\tilde{T}=T/n$.

Insertion into (\ref{eq:theproblem}) of
$\sum_{{\phi}^{\prime}}\delta_{{\phi_{i}}^{\alpha},{\phi}^{\prime}}=1$
followed by exponent linearization via Gaussian integrals gives
\begin{eqnarray*}
 f_N&=&-\frac{1}{\tilde{\beta} N}\log
\sum_{\bfeta} \sum_{\bphi^1\ldots\bphi^n}e^{ \frac{\beta
J}{N}\sum_{\phi^\prime}\sum_{\alpha=1}^n \left[\sum_i \eta_i
\delta_{\phi^\prime,\phi_i^\alpha}\right]^2 -\tilde{\beta}\sum_i
\mu_i\eta_i}
 \\
 &=&
 -\frac{1}{\tilde{\beta} N}\log
\int\![\prod_{\phi \alpha}dz_{\phi^\alpha}] e^{-\frac{N}{4\beta
J}\sum_{\phi\alpha}(z_\phi^\alpha)^2} \sum_{\bphi^1\ldots\bphi^n}
  \sum_{\bfeta}
e^{\sum_i \eta_i\left[ \sum_{\alpha} z^\alpha_{\phi_i^\alpha}
-\tilde{\beta}\mu_i\right]}
\end{eqnarray*}
We carry out the summations over the polarity variables (first)
and over the replicated angles, and  take the thermodynamic limit.
This leads us to the following result for the asymptotic free
energy per monomer $f=\lim_{N\to\infty}f_N$, which is evaluated by
steepest descent:
\begin{eqnarray}
\hspace*{-5mm} \tilde{\beta} f &=&
 - \lim_{N\to \infty}\frac{1}{ N}\log
\int\![\prod_{\phi \alpha}dz_\phi^\alpha]
 e^{-\frac{N}{4\beta
J}\sum_{\phi\alpha}(z_\phi^\alpha)^2+\sum_i \log \sum_{\bphi}
2\cosh\left[ \sum_{\alpha} z^\alpha_{\phi_\alpha}
-\tilde{\beta}\mu_i\right]} \nonumber\\
&=&
 {\rm extr}_{\{z_\phi^\alpha\}}\left\{
 \frac{1}{4\beta
J}\sum_{\phi\alpha}(z_\phi^\alpha)^2-\bra \log \left[
e^{-\tilde{\beta} \mu}\prod_\alpha (\sum_\phi
e^{z_\phi^\alpha})+e^{\tilde{\beta}\mu} \prod_\alpha (\sum_\phi
e^{-z_\phi^\alpha})\right] \ket_\mu \right\}
\label{eq:replicatedfenergy}
\end{eqnarray}
in which $\bphi=(\phi_1,\ldots,\phi_n)\in C^n$ and $\bra
g(\mu)\ket_\mu=\int\!d\mu~P(\mu)g(\mu)$, with
$P(\mu)=\lim_{N\to\infty}N^{-1}\sum_i \delta[\mu-\mu_i]$. The
location $\{Z_\phi^\alpha\}$ of the extremum in
(\ref{eq:replicatedfenergy}) is to be calculated by solving the
following non-linear saddle-point equations:
\be
Z_{\psi}^{\gamma}= 2\beta J\left\bra \frac{ e^{Z_\psi^\gamma-\beta
n \mu}\prod_{\alpha\neq \gamma} (\sum_\phi
e^{Z_\phi^\alpha})-e^{\beta n\mu-Z_\psi^\gamma} \prod_{\alpha\neq
\gamma} (\sum_\phi e^{-Z_\phi^\alpha})} { e^{-\beta n
\mu}\prod_\alpha (\sum_\phi e^{Z_\phi^\alpha})+e^{\beta n\mu}
\prod_\alpha (\sum_\phi e^{-Z_\phi^\alpha})} \right\ket_{\!\mu}
\label{eq:SPeqns} \ee The physical  meaning of the order
parameters $\{Z_\phi^\alpha\}$ can be determined by adding
generating terms to the Hamiltonian (\ref{eq:fastham}), which
measure the overall polarity  at given positions $\phi$:
 \be
 H(\bphi,\bfeta)\rightarrow
H(\bphi,\bfeta)+ \sum_{\phi}\chi_{\phi} L_{\phi}(\bphi,\bfeta)
~~~~~~~~ L_{\phi}(\bphi,\bfeta)=\frac{2}{N}\sum_{i=1}^N \eta_{i}
\delta_{\phi_{i},\phi} \label{eq:define_L}
 \ee
  Upon
working out the general identity $ \lim_{\chi_{\phi}\rightarrow
0}\partial f/\partial \chi_{\phi}= L_\phi=\overline{\bra
L_{\phi}({\bphi},{\bfeta}) \ket}$, with $\bra f({\bfeta,\bphi})
\ket$ denoting conformational equilibrium averages for fixed
$\{\eta_i\}$ and $\overline{f(\bfeta)}$ denoting  polarity
equilibrium averages, it follows that
 \be
 \frac{1}{n}\sum_\alpha Z_{\phi}^{\alpha}=\beta J L_\phi~~~~~~~~~~
 L_\phi= \lim_{N\to\infty}
\overline{\bra L_{\phi}({\bphi},{\bfeta}) \ket}
\label{eq:identify} \ee Thus $L_\phi$ is proportional to the
disorder-averaged equilibrium expectation value of the average
polarity of those monomers which are oriented to angle $\phi$. For
future use we will also define the overall average equilibrium
polarity $p$:
\be
p=\lim_{N\to\infty}\frac{1}{N}\sum_i
\overline{\eta_i}=\frac{1}{2}\sum_\phi L_\phi \label{eq:define_p}
\ee

\subsection{The replica symmetric solution}

In  \ref{app:stability} we show that the replica-symmetric (RS)
solution of our saddle-point equations (\ref{eq:SPeqns}), where
$Z_\phi^\alpha=\beta J L_\phi$ for all $\alpha$, is locally stable
against replica-symmetry breaking fluctuations. For such RS
solutions one finds the (\ref{eq:replicatedfenergy}) and
(\ref{eq:SPeqns}) reducing to, respectively:
\be
 f_{\rm RS} =
 {\rm min}_{\{\ell_\phi\}}\left\{
 \frac{J}{4}\sum_{\phi}\ell^2_\phi-\frac{1}{\beta n}\bra \log \left[(\sum_\phi
e^{\beta [J \ell_\phi- \mu]})^n +(\sum_\phi e^{\beta[\mu-J
\ell_\phi]})^n\right] \ket_\mu \right\} \label{eq:fRS} \ee
\be
 L_{\psi}= 2\left\bra \frac{ e^{\beta [J L_\psi- \mu]}(\sum_\phi e^{\beta [J L_\phi-\mu]})^{n-1}-e^{\beta[\mu-J L_\psi]}
(\sum_\phi e^{\beta[\mu-J L_\phi]})^{n-1} }{(\sum_\phi e^{\beta [J
L_\phi-\mu]})^{n}+ (\sum_\phi e^{\beta[\mu-
 J L_\phi]})^n}
\right\ket_{\!\mu} \label{eq:RSSPeqns} \ee
 Summation over $\phi$
in (\ref{eq:RSSPeqns}) gives an alternative expression for the RS
average polarity $p$ (\ref{eq:define_p}):
\be
p= \left\bra \frac{(\sum_\phi e^{\beta [J L_\phi-\mu]})^n-
(\sum_\phi e^{\beta[\mu-J L_\phi]})^n }{(\sum_\phi e^{\beta [J
L_\phi-\mu]})^{n}+ (\sum_\phi e^{\beta[\mu-
 J L_\phi]})^n}
\right\ket_{\!\mu} \label{eq:RSp}
 \ee
Equation (\ref{eq:RSp}) allows us to write (\ref{eq:RSSPeqns}) in
the form
\be
L_\psi=(1\plus p) \frac{e^{\beta J L_\psi}}{\sum_\phi e^{\beta
JL_\phi}} - (1\minus p)\frac{e^{-\beta J L_\psi}}{\sum_\phi
e^{-\beta JL_\phi}} \label{eq:RSSPeqnsnew} \ee This latter
expression (\ref{eq:RSSPeqnsnew}) was also found in
\cite{Skantzos}; however, here the average polarity $p$ is an
order parameter, to be solved simultaneously with the $\{L_\phi\}$
from (\ref{eq:RSp}), whereas in \cite{Skantzos} it was a fixed
control parameter.

 Since we have
shown replica symmetry to be locally stable, for any choice of
model parameters, we will henceforth restrict ourselves to RS
saddle-points only (there is no evidence for discontinuous RSB
transitions).

\section{General analytical results}

\subsection{The high-temperature state}

Let us first identify the high-temperature state. Expansion of the
free energy (\ref{eq:fRS}) for fixed $n>0$ gives $f_{\rm
RS}=-\frac{1}{2}\beta n\bra\mu^2\ket_\mu- \frac{1}{\beta}\log q
 -\frac{1}{\beta n}\log 2+ \frac{1}{2}J~{\rm extr}_{\{\ell_\phi\}} \Phi[\{\ell_\phi\}]+\order(\beta^2)$, with
(using $\sum_\phi 1=q$): \bd
 \Phi[\{\ell_\phi\}] =
 \frac{1}{2}\sum_{\phi}\ell^2_\phi
 +\frac{2\beta n}{ q}\bra\mu\ket_\mu\sum_\phi \ell_\phi
 -\beta J
\left[ \frac{1}{q} \sum_\phi  \ell^2_\phi  + (n\minus
1)(\frac{1}{q}\sum_\phi \ell_\phi)^2 \right] \ed It follows that
the saddle-point of $f$ is of the form \bd L_\phi=-\frac{2\beta
n}{q}\bra \mu\ket_\mu+\order(\beta^2)~~~~~~(\beta \to 0) \ed For
sufficiently high temperatures the order parameters $L_\phi$ are
thus  independent of $\phi$. This was to be expected in view of
the invariance of (\ref{eq:fRS}) under arbitrary permutations of
the available monomer orientations $\phi\in C$. \vsp

 Insertion of the symmetric ansatz $L_\phi=L$
$\forall\phi$ (which is a saddle-point of $f_{\rm RS}$ at any
temperature) directly into (\ref{eq:fRS},\ref{eq:RSSPeqns}) gives
\be
 f_{\rm RS}^{\rm sym} =
 {\rm min}_{\ell}\left\{
 \frac{Jq}{4}\ell^2
 -\frac{1}{\beta n}\bra \log \left[
e^{\beta n[J \ell- \mu]} + e^{\beta n[\mu-J \ell]} \right]
 \ket_\mu
\right\}
 -\frac{1}{\beta } \log q
 \label{eq:symmf}
 \ee
\be
 L= \frac{2}{q}\bra \tanh[\beta n(J L-\mu)]\ket_{\mu}
\label{eq:symmstate}
 \ee
In terms of the average equilibrium polarity
 (\ref{eq:define_p}), which here reduces to
 $p=qL/2$, equation (\ref{eq:symmstate}) can be written alternatively as
$p= \bra \tanh[\beta n(2Jp/q-\mu)]\ket_{\mu}$. The instance where
this RS high-temperature state becomes locally unstable can be
inferred from the results of \ref{app:stability}. In particular
(\ref{eq:RSeigenvalueidentity}) gives the condition for a zero
eigenvalue of the RS Hessian as \bd {\rm det}
\left[nq(\hat{\bK}-\hat{\bL})+q(\hat{\bL}-\hat{\bM})+\frac{q}{2\beta
J}\unity \right]=0 \ed Working out the various matrices for the
symmetric state $L_\phi=L$, with (\ref{eq:symmstate}), gives \bd
\hat{K}_{\phi\psi}= q^{-2}\bra
 \tanh^2[n\beta( J L\minus\mu)]
 \ket_\mu
~~~~~~~~~~
 \hat{L}_{\phi\psi}= q^{-2}
~~~~~~~~~~ \hat{M}_{\phi\psi}=q^{-1}\delta_{\phi\psi} \ed In the
symmetric state the Hessian has two distinct eigenvalues, one
relating to changes in the amplitude of the symmetric state, and
one relating to changes orthogonal to the symmetric state, with
associated stability conditions:
 \begin{eqnarray*}
{\rm locally~stable~in~non\!\!-\!\!symmetric~directions:}&~~~&
qT/2 J>1\\ {\rm locally~stable~in~symmetric~direction:} &~~~& qT/2
J> n\bra 1\minus
 \tanh^2[n\beta( J L\minus \mu)]\ket_\mu
 \end{eqnarray*}
 For $n\leq 1$ it immediately follows that, as we lower the temperature, the first stability
 condition is always violated before the second can be. Violation of the second condition implies destabilization in
favour of an alternative symmetric solution (i.e. the creation of
an alternative solution of equation (\ref{eq:symmstate})). Thus
symmetric states $L_\phi=L$ $\forall\phi$ defined by
(\ref{eq:symmf}) become locally unstable against symmetry-breaking
fluctuations at
\be
T_c=2J/q \label{eq:criticaltemp} \ee Unless preceded by a first
order transition (which will also turn out to be possible in
certain parameter regimes), this describes a second-order phase
transition from a non-separated state into one where the
hydrophobic and hydrophilic monomers start to prefer distinct
orientations. Within the context of our model, the former is to be
regarded as a swollen state (S) for the polymer, and the second as
a compact state (C).  Note that expression
(\ref{eq:criticaltemp}), which is identical to that found in
\cite{Skantzos}, is independent of $n$.

\subsection{The replica-symmetric ground state for $n>0$}

We now study the limit $\beta\to\infty$ and calculate the ground
state of our system, for replica-symmetric solutions and fixed
$n>0$. We define $\ell_+=\max_\phi \ell_\phi$, $\ell_-=\min_\phi
\ell_\phi$ (so $\ell_{+}\geq \ell_{-}$) and the number of
locations $\phi$ for which $\ell_{\phi}=\ell_{\pm}$ by $q_{\pm}$,
respectively (so $q\pm\geq 1$). The remaining intermediate values
for $\ell_\phi$ define the set
$U=\{\phi|~\ell_-<\ell_\phi<\ell_+\}$. Similarly we define $L_\pm$
as the  values of $\ell_\pm$ at the relevant saddle-point. The
ground state is the solution of the following minimization
problem:
\begin{eqnarray*}
\hspace*{-7mm}
 \frac{E_0}{J} &=&
 {\rm min}_{\{\ell_\phi\}}\left\{
 \frac{1}{4}\sum_{\phi}\!\ell^2_\phi-\lim_{\beta\to\infty}\frac{1}{\beta Jn}\bra ~\log \left[(\sum_\phi
e^{\beta [J \ell_\phi- \mu]})^n +(\sum_\phi e^{\beta[\mu-J
\ell_\phi]})^n\right] \ket_\mu \right\}
\\
 &=&
 {\rm min}_{\{\ell_\phi\}}\left\{
 \frac{1}{4}\sum_{\phi}\!\ell^2_\phi-\frac{1}{J}\bra \max\left[
  \lim_{\beta\to\infty}\frac{1}{\beta}\log (\sum_\phi
e^{\beta [J \ell_\phi- \mu]}),\lim_{\beta\to\infty}\frac{1}{\beta}
\log (\sum_\phi e^{\beta[\mu-J \ell_\phi]})\right] \ket_\mu
\right\}
\\
 &=&
 {\rm min}_{\{\ell_\phi\}}\left\{
\frac{1}{4}\sum_{\phi}\!\ell^2_\phi-\bra~ \max\left[
 \ell_+ \!- \mu/J,~\mu/J- \ell_-\right]~ \ket_\mu \right\}
\end{eqnarray*}
Suppose first that $L_+=L_-$, which implies $L_\phi=L=2p/q$ for
all $\phi$ (i.e. a fully symmetric ground state). Now
\begin{eqnarray*}
 E_0/J &=&
  {\rm min}_{\{p\}}\left\{
 p^2/q -\bra~
 |2p/q - \frac{\mu}{J}|~ \ket_\mu
 \right\}
\end{eqnarray*}
Next we consider the case $L_+>L_-$. Minimization with respect to
those $\ell_\phi$ for which $\phi\in U$ for a given realization of
$\{q_+,q_-\}$ reveals that $\ell_\phi=0$ for $\phi\in U$. We can
then minimize further with respect to $\{q_\pm\}$ and find
$q_+=q_-=1$. Given our previous identification
$p=\frac{1}{2}\sum_\phi L_\phi$, here reducing to
$p=\frac{1}{2}(L_+\plus L_-)$, we write $\ell_\pm=p\pm z$, upon
which our problem takes the form
\begin{eqnarray*}
 E_0/J &=&
  {\rm min}_{\{p,z\}}\left\{
\frac{1}{2} p^2 +\frac{1}{2}z^2-z
 - \bra~|p- \mu/J|~\ket_\mu
 \right\}
\end{eqnarray*}
We conclude that the minimum corresponds to $x=1$. Both candidate
ground state solutions can thus be expressed in terms of the
monotonically increasing function \bd \Lambda(z)=
  {\rm min}_{\{y\}}\left\{
\frac{1}{2}z y^2  - \bra |y- \mu/J| \ket_\mu\right\}
 \ed
  as follows
\begin{eqnarray*}
{\rm symmetric:} ~~~~~~~~~~~ \bL=L(1,1,\ldots,1,1), &~~~&
 E_0/J = \Lambda(q/2)
 \\
{\rm non\!-\!symmetric:}~~~ \bL=(p\minus 1,0,\ldots,0,p\plus
1),&~~~&
 E_0/J = \Lambda(1)-1/2
\end{eqnarray*}
Since $q\geq 2$ we may conclude that the ground state is the
non-symmetric solution, so
\be
 E_0/J =
  {\rm min}_{\{p\}}\epsilon(p)
  ~~~~~~~~
  \epsilon(p)=
\frac{1}{2} p^2 -\frac{1}{2} - \bra~|p- \mu/J|~\ket_\mu
\label{eq:groundstate_nonsym}
 \ee
 Note that $\epsilon(p)$ is symmetric as soon as $P(\mu)$ is symmetric. The
ground state is one where at most two sites $\phi_\pm$ are
occupied by monomers, and there are at most two nonzero values
$L_\pm$ for the order parameters $L_\phi$.

\subsection{The limits $n\to 0$, $n\to 1$ and $n\to\infty$, for arbitrary $T$}

If we take $n\to 0$ in equations
(\ref{eq:fRS},\ref{eq:RSSPeqns},\ref{eq:RSp}) (which implies fully
random evolution of the polarity variables $\eta_i$) we find, as
expected, the $p=0$ version of the solution for the long-range
limit of \cite{Skantzos}:
 \bd \lim_{n\to 0}\left[ f_{\rm RS}
\plus\frac{\log 2}{\beta n}\right]
=
 {\rm min}_{\{\ell_\phi\}}\left\{
 \frac{J}{4}\sum_{\phi}\ell^2_\phi-\frac{1}{2\beta}
 \log(\sum_\phi e^{\beta J \ell_\phi})
-\frac{1}{2\beta}
 \log (\sum_\phi e^{-\beta J \ell_\phi})
  \right\}
 \ed
\bd
 L_{\psi}= \frac{e^{\beta J L_\psi}}{\sum_\phi e^{\beta J L_\phi}}
 -\frac{e^{-\beta J L_\psi}}{
\sum_\phi e^{-\beta J L_\phi}},~~~~~~~~~~p=0  \ed The (constant)
contribution $-\log 2/\beta n$ to the free energy per monomer
represents the entropy of the polarity variables. Note that this
solution is independent of the distribution $P(\mu)$ of the forces
$\mu_i$, as it should. \vsp

Putting $n\to 1$ effectively reduces the polarity variables to
annealed ones, in spite of the disparate time-scales in the model.
Now we find (\ref{eq:fRS},\ref{eq:RSp},\ref{eq:RSSPeqnsnew})
taking the form \bd \lim_{n\to 1} f_{\rm RS} =
 {\rm min}_{\{\ell_\phi\}}\left\{
 \frac{J}{4}\sum_{\phi}\ell^2_\phi-\frac{1}{\beta }\bra \log
 2\sum_\phi \cosh[\beta (J \ell_\phi\minus \mu)]\ket_\mu \right\}
 \ed \bd
 L_\psi=(1\plus p) \frac{e^{\beta J L_\psi}}{\sum_\phi e^{\beta
JL_\phi}} - (1\minus p)\frac{e^{-\beta J L_\psi}}{\sum_\phi
e^{-\beta JL_\phi}},
 ~~~~~~~~~~ p= \bigbra \frac{\sum_\phi \sinh[\beta(J L_\phi\minus
\mu)]}{\sum_\phi \cosh[\beta (J L_\phi\minus \mu)]}
\bigket_{\!\mu} \ed Somewhat unexpectedly, these  equations are
still non-trivial, and retain a rich bifurcation phenomenology as
we will show later for specific choices of the number $q$ of
possible monomer orientations. \vsp

Now we turn to $n\to\infty$, i.e. to fully deterministic genetic
dynamics. Here the free energy per monomer (\ref{eq:fRS}) reduces
to
\begin{eqnarray}
\hspace*{-10mm} \lim_{n\to\infty} f_{\rm RS}& =&
 {\rm min}_{\{\ell_\phi\}}\left\{
 \frac{J}{4}\sum_{\phi}\ell^2_\phi-\bra
\max\left[
 \frac{1}{\beta}\log(\sum_\phi e^{\beta J \ell_\phi})\minus\mu,\mu\plus\frac{1}{\beta}\log(\sum_\phi
e^{-\beta J \ell_\phi})\right] \ket_\mu \!\right\} \nonumber
\\
& =&
 {\rm min}_{\{\ell_\phi\}}\left\{
 \frac{J}{4}\sum_{\phi}\ell^2_\phi
 -
 \frac{1}{2\beta}\log(\sum_\phi e^{\beta J
\ell_\phi}) -
 \frac{1}{2\beta}\log(\sum_\phi e^{-\beta J \ell_\phi}) \right.
\nonumber\\ &&\left. \hspace*{30mm}
 -\bigbra
\left|\mu-\frac{1}{2\beta}\log\left[\frac{\sum_\phi e^{\beta J
\ell_\phi}}{\sum_\phi e^{-\beta J \ell_\phi}}\right]\right|
\bigket_{\!\!\mu} \right\}
\end{eqnarray}
Equation (\ref{eq:RSp}), similarly, becomes
\begin{eqnarray}
p&=&
-\bigbra\!\sgn\left[\mu-\frac{1}{2\beta}\log\left[\frac{\sum_\phi
e^{\beta J L_\phi}}{\sum_\phi e^{-\beta J L_\phi}}\right]\right]
\bigket_{\!\!\mu}
\end{eqnarray}

\subsection{Phase diagrams for $q=2$  and $q=3$}

In view of our earlier results regarding the high-temperature
states (where $\bL =L(1,1,\ldots,1,1)$) and ground states (where
$\bL=(p-1,0,\ldots,0,p+1)$) it is natural to assume that the
solution of our model will never exhibit more than three distinct
values for the order parameters $\{L_\phi\}$. Hence we may
restrict our analysis to $q\in\{2,3\}$ and expect at most
quantitative changes to emerge for $q>3$. In the phase diagrams to
be given below for $q\in\{2,3\}$ we focus on transitions marking
bifurcations of local minima of the free energy surface as
minimized in (\ref{eq:fRS}), rather than on thermodynamic
transitions. Since all energetic and entropic barriers in our
system are extensive, on the time-scales of any experiment
(whether real or numerical, and especially in view of the
extremely slow genetic process) one would never detect
thermodynamic transitions. The system will in practice be found in
the local free energy minimum in whose domain of attraction the
initial configuration happened to lie.

 For $q=2$ we have only two order parameters, which we can
write as $L_\pm=p\pm Z$. Insertion into (\ref{eq:fRS}) reveals
that now the free energy minimization decouples conveniently into
\be
\hspace*{-20mm}
 f_{\rm RS} =
 {\rm min}_{\{Z\}}\left\{
 \frac{JZ
 z^2}{2}-\frac{1}{\beta } \log 2\cosh[\beta JZ] \right\}
  +
 {\rm min}_{\{p\}}\left\{
 \frac{Jp^2}{2}
-\frac{1}{\tilde{\beta}}\bra \log 2\cosh[\tilde{\beta}(Jp\minus
\mu)] \ket_\mu \right\} \label{eq:f_q=2} \ee giving the following
independent order parameter equations:
\be
 Z= \tanh[\beta JZ] ~~~~~~~~~~
 p=\bra \tanh[\tilde{\beta}(Jp-\mu)]
\ket_\mu \label{eq:saddle_q2} \ee The separation process of
hydrophobic from hydrophilic monomers (i.e. the folding), measured
 by $Z$, disentangles from the species evolution, as measured by
$p$; with the two processes each having their own independent
transitions. This is a consequence of the fact that for $q=2$ the
orientation variables $\phi_i$ effectively become Ising spins, so
that upon putting $\delta_{\phi_i\phi_j}\to
\frac{1}{2}[1+\sigma_i\sigma_j]$, the `folding' Hamiltonian in
(\ref{eq:theproblem}) reduces to that of a Mattis \cite{Mattis}
magnet, from which the variables $\{\eta_i\}$ can be gauged away.

The uniform high-temperature state $L_\phi=L$ $\forall\phi$ (where
$Z=0$) always destabilizes via a second order phase transition at
the Curie-Weiss temperature $T_c=J$, independent of the
distribution $P(\mu)$.
 The phase phenomenology embodied in the
equation for $p$, in contrast, is described by an equation for a
mean-field ferro-magnet at inverse temperature $\tilde{\beta}$ and
with random external fields, distributed (apart from a minus sign)
according to $P(\mu)$. It will be therefore be dependent on the
choice made for $P(\mu)$, with invariance of the problem under
$p\to -p$ for symmetric force distributions $P(\mu)$.

For $q=3$ one has $\phi\in \{ -\frac{2}{3}\pi,0,\frac{2}{3}\pi
\}$, and there will be three order parameters $\{L_\phi\}$. One
can no longer map the model onto a Mattis-like system, and one no
longer benefits from  the resulting decoupling of orientation
degrees of freedom from polarity degrees of freedom which occurred
for $q=2$.
 According to (\ref{eq:criticaltemp}), the fully
symmetric (i.e. swollen) state $L_\phi=L$ $\forall\phi$ now
destabilizes locally at the $n$-independent temperature
$T_c=2J/3$. In addition we know the ground state of the $q=3$
system, for several choices of the force distribution $P(\mu)$,
see e.g. (\ref{eq:groundstate_delta},\ref{eq:groundstate_binary}).
 In contrast to the previous situation $q=2$, however, there appear to be no
significant\footnote{It will be clear that our equations can still
be simplified partially upon making specific {\em ans\"{a}tze} for
the saddle-point as in \cite{Skantzos}, such as the uniform (i.e.
swollen) state $L_\phi=L$ $\forall\phi$ (which has already been
studied in detail for arbitrary $q$), or states of the form
$\bL=(L_1,L_2,L_2)$ (where our saddle-point and stability problems
become 2-dimensional). Note that our system is invariant under
permutations of the three monomer orientations, so that states
like $\bL=(L_1,L_2,L_2)$ and $\bL=(L_2,L_2,L_1)$ are equivalent.}
further analytical simplifications possible, and to obtain results
on compact states and phase transitions  in the regime of
intermediate temperatures we have to resort mainly to a numerical
analysis of our fundamental equations
(\ref{eq:fRS},\ref{eq:RSSPeqns},\ref{eq:RSp},\ref{eq:RSSPeqnsnew}).
Bifurcations of new solutions of the saddle-point equations are
marked by the smallest eigenvalue of the relevant Hessian becoming
zero, i.e.
\be
\left[n(\hat{\bK}\minus
\hat{\bL})+\hat{\bL}-\hat{\bM}\right]\bx=-\frac{1}{2} \beta J \bx
\label{eq:q=3transitions} \ee (see \ref{app:stability}), with the
$3\times 3$ matrices as defined in
(\ref{eq:hatK},\ref{eq:hatL},\ref{eq:hatM}). Extensive numerical
analysis of the phases in the region $T<2J/3$, where the swollen
state is locally unstable, reveals a highly non-trivial phase
phenomenology, with many simultaneously locally stable
saddle-points. Giving all details and lines in this regime of
compact states  would be more distracting than informative. In
contrast, we will focus mainly on transitions in the region
$T>2J/3$, where the swollen state is locally stable, but where in
spite of this one generally finds enhanced first order transitions
to compact states whose presence is a direct consequence of our
coupled dynamics of slow and fast processes\footnote{In the
long-range limit of \cite{Skantzos} one can also find
discontinuous bifurcations for very specific values of the control
parameters (marking the creation a locally stable state with
relatively high free energy, and hence non-thermodynamic), and
very close to the general second order transition at $T_c=2J/q$;
this was missed in \cite{Skantzos}.}. Note that for $q=3$ we give
phase diagrams with $\beta J$ along the horizontal axis (as
opposed to $\tilde{\beta}J=n\beta J $) because, in contrast to
$q=2$, one here no longer finds that the non-trivial transitions
can be expressed in terms of just two effective control parameters
($\beta nJ$ and $\beta n \sigma$ or $\beta n J$).

\section{Results for $\delta$-distributed genetic forces}

In the simplest case $P(\mu)=\delta[\mu-\overline{\mu}]$ (i.e.
$\mu_i=\overline{\mu}$ for all $i$) the functionality potential
$V(\bfeta)$ in the genetic dynamics favors a single polarity type.
Since the general dynamical behaviour of our system can be
regarded as a stochastic version of the combined $T=0$ (i.e.
deterministic) folding process and the $n\to\infty$ (i.e.
deterministic) genetic dynamics, we first inspect the limits $T\to
0$ and $n\to\infty$.

\subsection{The deterministic limits: $T\to 0$ and $n\to\infty$}

We first calculate the ground state (i.e. deterministic evolution
of configurational angles, $T=0$). We find the function
$\epsilon(p)$ in (\ref{eq:groundstate_nonsym}) reducing to \bd
  \epsilon(p)=
\frac{1}{2} p^2 -\frac{1}{2} - |p\!- \frac{\overline{\mu}}{J}| \ed
Upon working out the derivatives of $\epsilon(p)$ in the different
regimes (characterized by different values of
$\sgn[p-\overline{\mu}/J]$)
 one  arrives at the following result:
\be
E_0=- J-
|\overline{\mu}|,~~~~~~~~\bL=2p(1,0,\ldots,0),~~~~~~~~\left\{\begin{array}{lllll}
\overline{\mu}>0:&& p=-1\\ \overline{\mu}<0:&& p=1
\end{array} \right.
\label{eq:groundstate_delta}
 \ee
For $\overline{\mu}>0$ the ground state is one where {\em all}
monomers have become hydrophobic, and are found at exactly the
same location relative to the chain. For $\overline{\mu}<0$ the
ground state is one where {\em all} monomers have become
hydrophilic, and are again all oriented at the same location.
However, one finds also that for $|\overline{\mu}|<J$ the
solutions $p=\pm 1$ are both local minima of $\epsilon(p)$
(although the state $p=\sgn[\overline{\mu}]$ will have an energy
higher than the ground state $p=-\sgn[\overline{\mu}]$ as long as
$\overline{\mu}\neq 0$). This will cause remanence effects in the
low temperature dynamics. For $\overline{\mu}=0$ both
single-species states $p=\pm 1$ give equivalent (local and global)
minima.

Next we turn to $n\to\infty$, i.e. deterministic genetic dynamics.
Equation (\ref{eq:RSp}), together with the order-parameter
equation (\ref{eq:RSSPeqnsnew}), give us three candidate solution
classes, $p\in\{-1,0,1\}$ for $n\to\infty$. Only $p=\pm 1$ will be
potentially stable, separated by the unstable fixed-point $p=0$:
\begin{eqnarray*}
p=1:~~~&&  L_\psi= \frac{ 2 e^{\beta J L_\psi}}{\sum_\phi e^{\beta
JL_\phi}},~~~~~~~~~~~~~~~~~~~~~~\overline{\mu}<\frac{1}{2\beta}\log\left[\frac{\sum_\phi
e^{\beta J L_\phi}}{\sum_\phi e^{-\beta J L_\phi}}\right]
\nonumber
\\
&& \lim_{n\to\infty} f_{\rm RS} =
 \frac{J}{4}\sum_{\phi}L^2_\phi
 -
 \frac{1}{\beta}\log(\sum_\phi e^{\beta J
L_\phi})
\\
 p=-1:~~~&& L_\psi= \frac{ -2 e^{-\beta J
L_\psi}}{\sum_\phi e^{-\beta
JL_\phi}},~~~~~~~~~~~~~~~~~~~~~\overline{\mu}>\frac{1}{2\beta}\log\left[\frac{\sum_\phi
e^{\beta J L_\phi}}{\sum_\phi e^{-\beta J L_\phi}}\right]
\nonumber \\ &&
 \lim_{n\to\infty} f_{\rm RS} =
 \frac{J}{4}\sum_{\phi}L^2_\phi -
 \frac{1}{\beta}\log(\sum_\phi e^{-\beta J L_\phi})
 \end{eqnarray*}
For the swollen state $L_\phi=2p/q$ $\forall \phi$ this implies
the following: for $|\overline{\mu}|>2J/q$ there is only the
solution $p=-\sgn[\overline{\mu}]$, for $|\overline{\mu}|<2J/q$
both solutions $p=\pm 1$ are locally stable (but with the lowest
free energy obtained for $p=-\sgn[\overline{\mu}]$).

In both cases the system evolves towards a single-species state,
with either all hydrophilic monomers (the preferred option when
$\overline{\mu}<0$) or all hydrophobic monomers (the preferred
option when $\overline{\mu}>0$). For  small $|\overline{\mu}|$
only the preferred states are locally stable, but for sufficiently
large $|\overline{\mu}|$ both single-species states are. This
behaviour, which is not desirable from a biological point of view,
results from the simple fact that for
$P(\mu)=\delta[\mu-\overline{\mu}]$ the single-species state with
appropriate polarity sign is energetically favorable to both the
folding process and the genetic selection process.

\subsection{Phase diagrams for $q=2$}

\begin{figure}[t]
\vspace*{18mm}\hspace*{1mm} \setlength{\unitlength}{0.62mm}
\begin{picture}(0,70)
\put(0,-10){\epsfxsize=140\unitlength\epsfbox{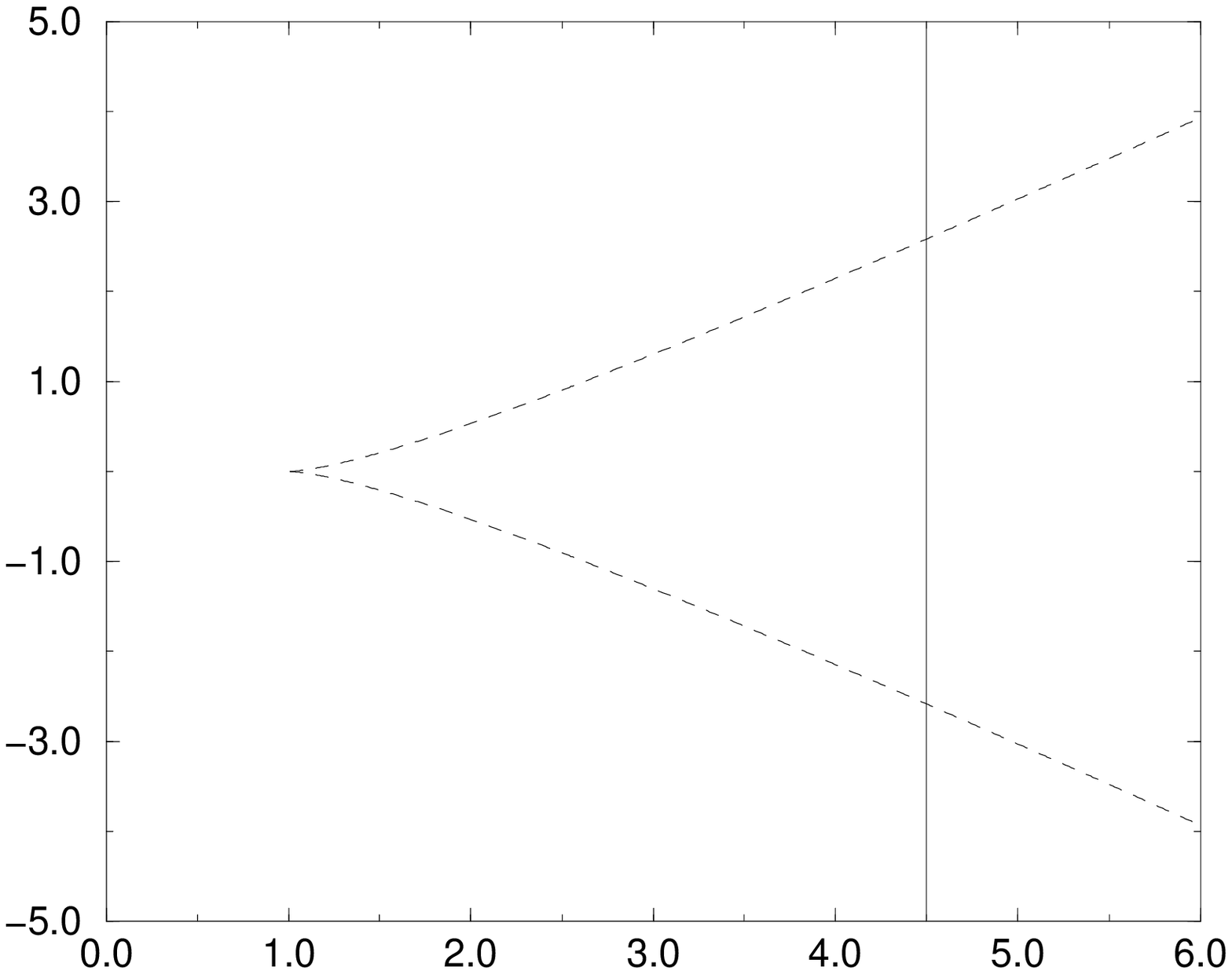}}
\put(0, 50){\large $\tilde{\beta}\overline{\mu}$}
\put(69,-9){\large $\tilde{\beta} J$}
\put(140,-10){\epsfxsize=140\unitlength\epsfbox{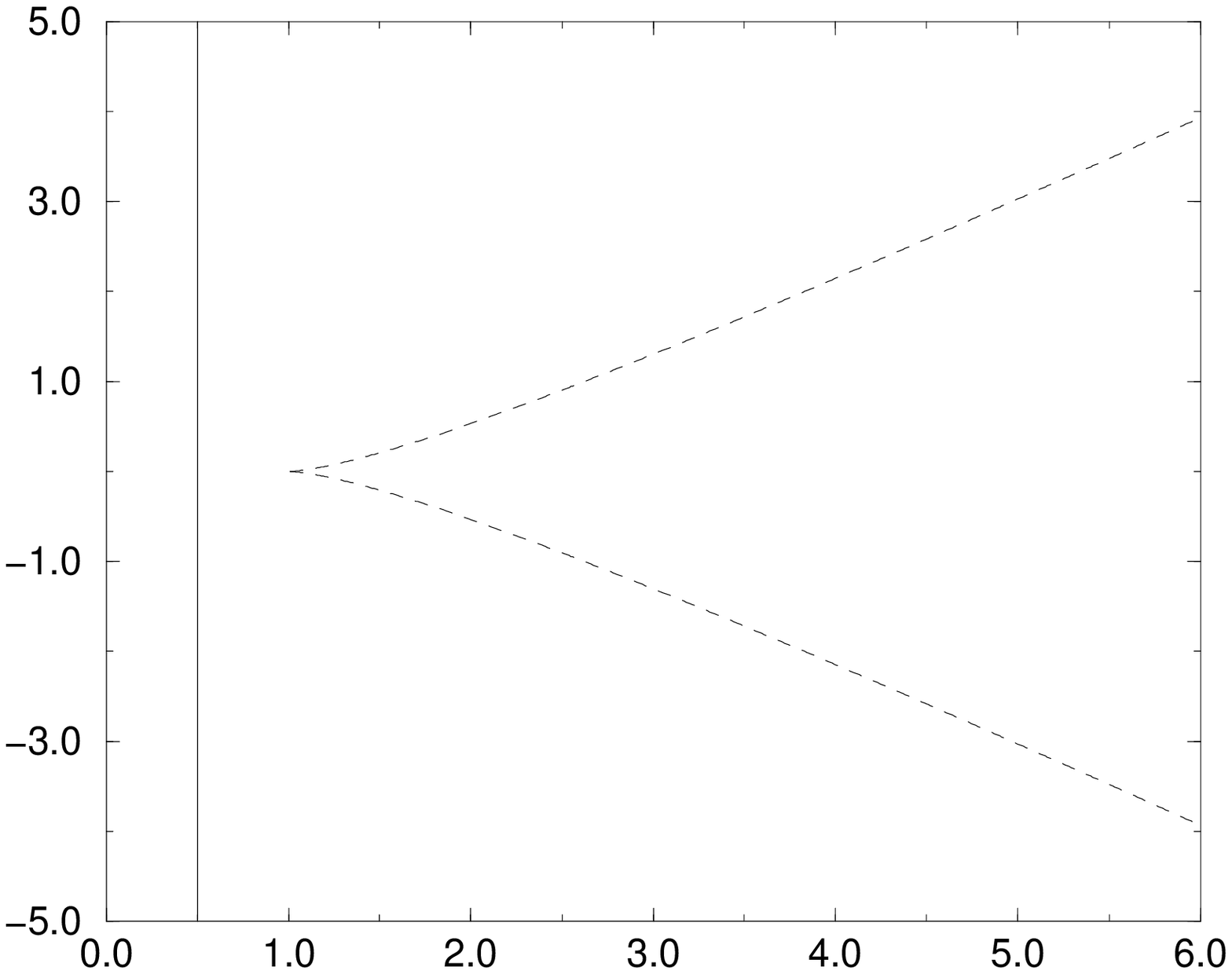}}
\put(140,  50){\large $\tilde{\beta}\overline{\mu}$}
\put(209,-9){\large $\tilde{\beta} J$} \put( 50,  80){$S1$} \put(
108, 85){$C1$} \put( 80, 50){$S2$} \put( 108, 50){$C2$}
\put(164,50){$S1$} \put( 50, 20){$S1$} \put( 108,  15){$C1$} \put(
200, 80){$C1$} \put( 230, 50){$C2$} \put( 200, 20){$C1$}
\end{picture}
\vspace*{10mm} \caption{Phase diagrams for $q=2$ and
$P(\mu)=\delta[\mu-\overline{\mu}]$. The four possible phases are
$\{S1,S2\}$ (swollen phases with $1$ or $2$ possible locally
stable values of $p$)  and $\{C1,C2\}$ (compact phases with $1$ or
$2$ possible locally stable values of $p$). The $S1\to C1$ and
$S2\to C2$   transitions (at $\tilde{\beta}J=n$) are second order
(they mark the creation of $Z\neq 0$ solutions of
(\ref{eq:saddle_q2})). The $S1\to S2$ and $C1\to C2$ transitions
are first order (except at $\overline{\mu}=0$, where the
transition is second order). Left diagram: $n=5$. In this and
subsequent phase diagrams all first and second order transitions
are drawn as dashed and solid curves, respectively. Right diagram:
$n=\frac{1}{2}$. Note that $\tilde{\beta}=\beta n$ and that phase
$S2$ exists only for $n>1$.} \label{fig:singlemuphasediagramq=2}
\end{figure}

According to (\ref{eq:saddle_q2}),  for
$P(\mu)=\delta[\mu-\overline{\mu}]$ we find the simple Curie-Weiss
equation $ p=\tanh[\tilde{\beta}(Jp-\overline{\mu})]$. Upon
inverting this to
$\tilde{\beta}\overline{\mu}=\tilde{\beta}Jp-\frac{1}{2}\log[(1\plus
p)/(1\minus p)]$ and calculating the derivative with respect to
$p$, one can work out the bifurcation properties of the solution:
\begin{eqnarray*}
\tilde{\beta}J
>1~~{\rm and}~~|\overline{\mu}|<\mu_c: &~~~& {\rm two~local~minima~}p~~{\rm
in~(\ref{eq:f_q=2})}\\ {\rm elsewhere:} &~~&  {\rm
one~local~minimum~}
\end{eqnarray*}
with
\be
\tilde{\beta}\mu_c=\sqrt{\tilde{\beta}J}\sqrt{\tilde{\beta}J
\minus 1} -\frac{1}{2} \log\left[ \frac{\sqrt{\tilde{\beta}J}\plus
\sqrt{\tilde{\beta}J \minus 1}}{\sqrt{\tilde{\beta}J}\minus
\sqrt{\tilde{\beta}J\minus 1}} \right]
 \label{eq:phaseequation}
\ee The transitions at $\mu=\pm \mu_c$ are first-order, except for
the common point
$(\tilde{\beta}J,\tilde{\beta}\overline{\mu})=(1,0)$ where they
become second-order. In the region of multiple local minima, for
$\overline{\mu}\neq 0$ the global minimum has
$\sgn[p]=-\sgn[\overline{\mu}]$, whereas for $\overline{\mu}=0$
the local minima for $\tilde{\beta}J>1$ are equivalent. We note
\begin{eqnarray*}
\tilde{\beta}\mu_c=\frac{1}{2}(\tilde{\beta}J\minus 1)^{3/2}
+\order((\tilde{\beta}J\minus 1)^2) &~~~~~& (\tilde{\beta}J\to
1)\\
\tilde{\beta}\mu_c=\tilde{\beta}J+\order(\log(\tilde{\beta}J)) &
~~~~~& (\tilde{\beta}J\to \infty)
\end{eqnarray*}
This agrees with our results regarding the ground state. We have
now determined all phases and transitions: there is one second
order transition at $\beta J=1$ from a swollen state with
uniformly distributed monomer orientations  to a compact state
with separation of polarity types, and two first-order transitions
at $\overline{\mu}=\pm \mu_c(\beta,n,J)$ (in the region $\beta n
J>1$) marking the creation of multiple locally stable values for
the average polarity. These lines are shown in the
$(\tilde{\beta}J,\tilde{\beta}\overline{\mu})$ phase diagram, in
figure \ref{fig:singlemuphasediagramq=2}. We denote swollen phases
with $\ell$ possible locally stable values of $p$ as $S_\ell$, and
compact phases with $\ell$ possible locally stable values of $p$
as $C_\ell$. The number of possible phases depends explicitly on
the value of $n$, since phase $S2$ exists only for $n>1$.

\subsection{Phase diagrams for $q=3$}

\begin{figure}[t]
\vspace*{18mm}\hspace*{1mm} \setlength{\unitlength}{0.62mm}
\begin{picture}(0,70)
\put(0,-10){\epsfxsize=140\unitlength\epsfbox{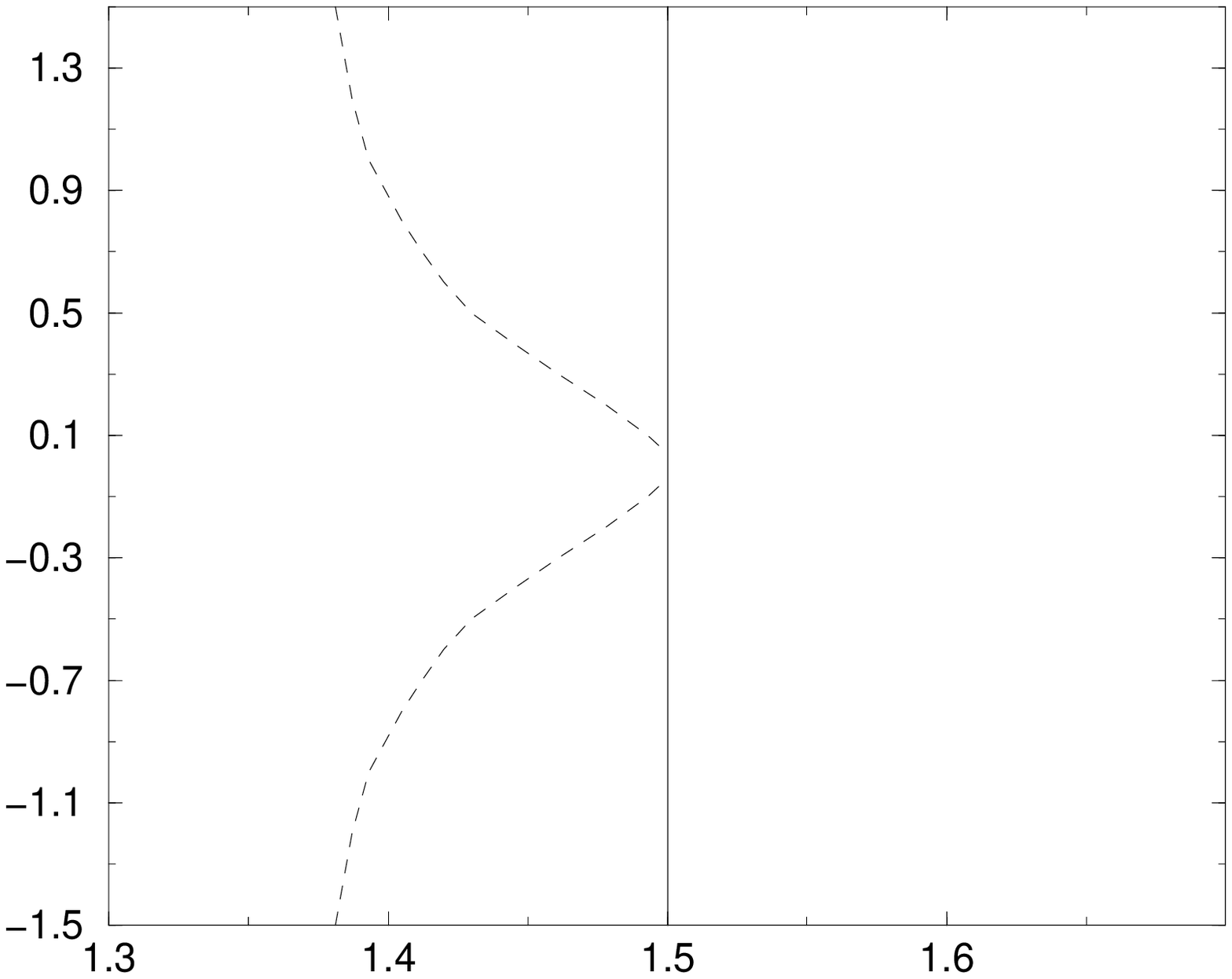}}
\put(0, 50){\large $\tilde{\beta}\overline{\mu}$}
\put(69,-9){\large $\beta J$}
 \put( 50,  80){$S1\!+\!C1$}  \put(95, 50){$CO$}
 \put(40, 50){$S1$} \put(50,20){$S1\!+\!C1$}
\put(140,-10){\epsfxsize=140\unitlength\epsfbox{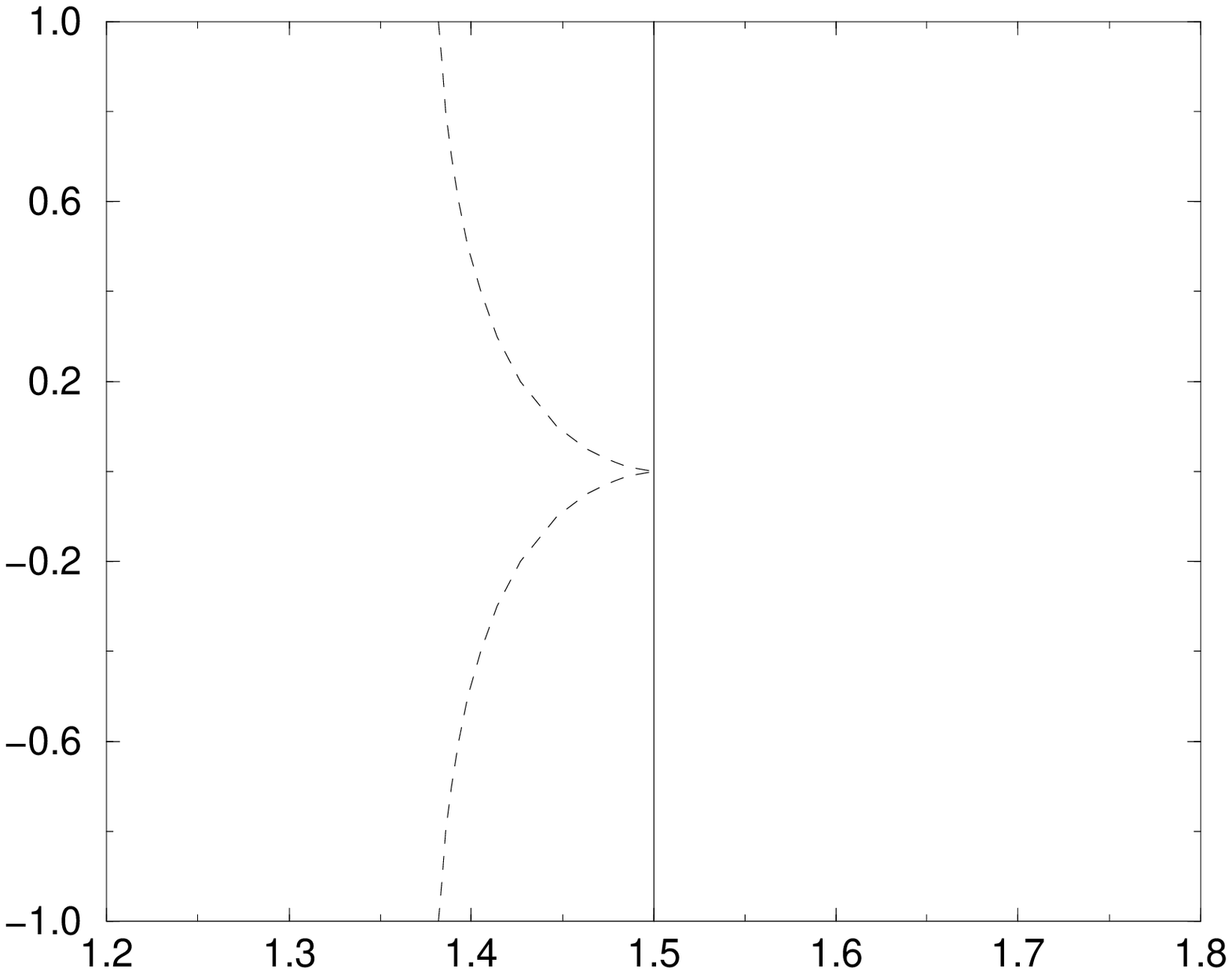}}
\put(140,  50){\large $\tilde{\beta}\overline{\mu}$}
\put(209,-9){\large $\beta J$}
 \put( 195,  87){$S1\!+\!C1$}  \put(235, 50){$CO$}
 \put(180, 50){$S1$} \put(195,13){$S1\!+\!C1$}
\end{picture}

\vspace*{28mm}\hspace*{1mm}
\begin{picture}(0,70)
\put(0,-10){\epsfxsize=140\unitlength\epsfbox{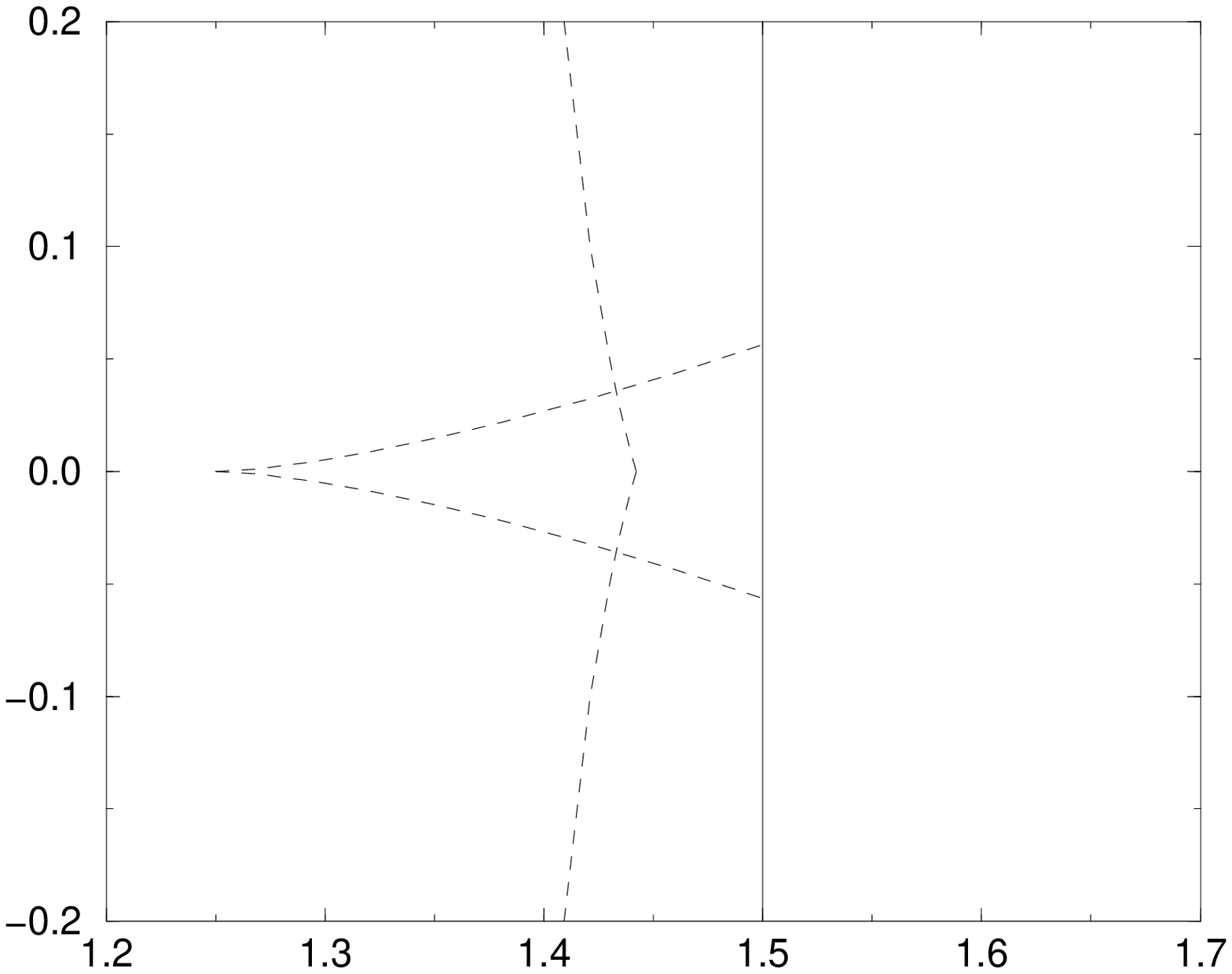}}
\put(0, 50){\large $\tilde{\beta}\overline{\mu}$}
\put(69,-9){\large $\beta J$}
 \put(66,  87){$S1\!+\!C1$}  \put(100, 50){$CO$}
 \put(40, 75){$S1$}  \put(40, 25){$S1$}
 \put(60, 50){$S2$}
 \put(76,56){$S2$}\put(76,50){$+$}\put(76,44){$C2$}
 \put(66,13){$S1\!+\!C1$}
\put(140,-10){\epsfxsize=140\unitlength\epsfbox{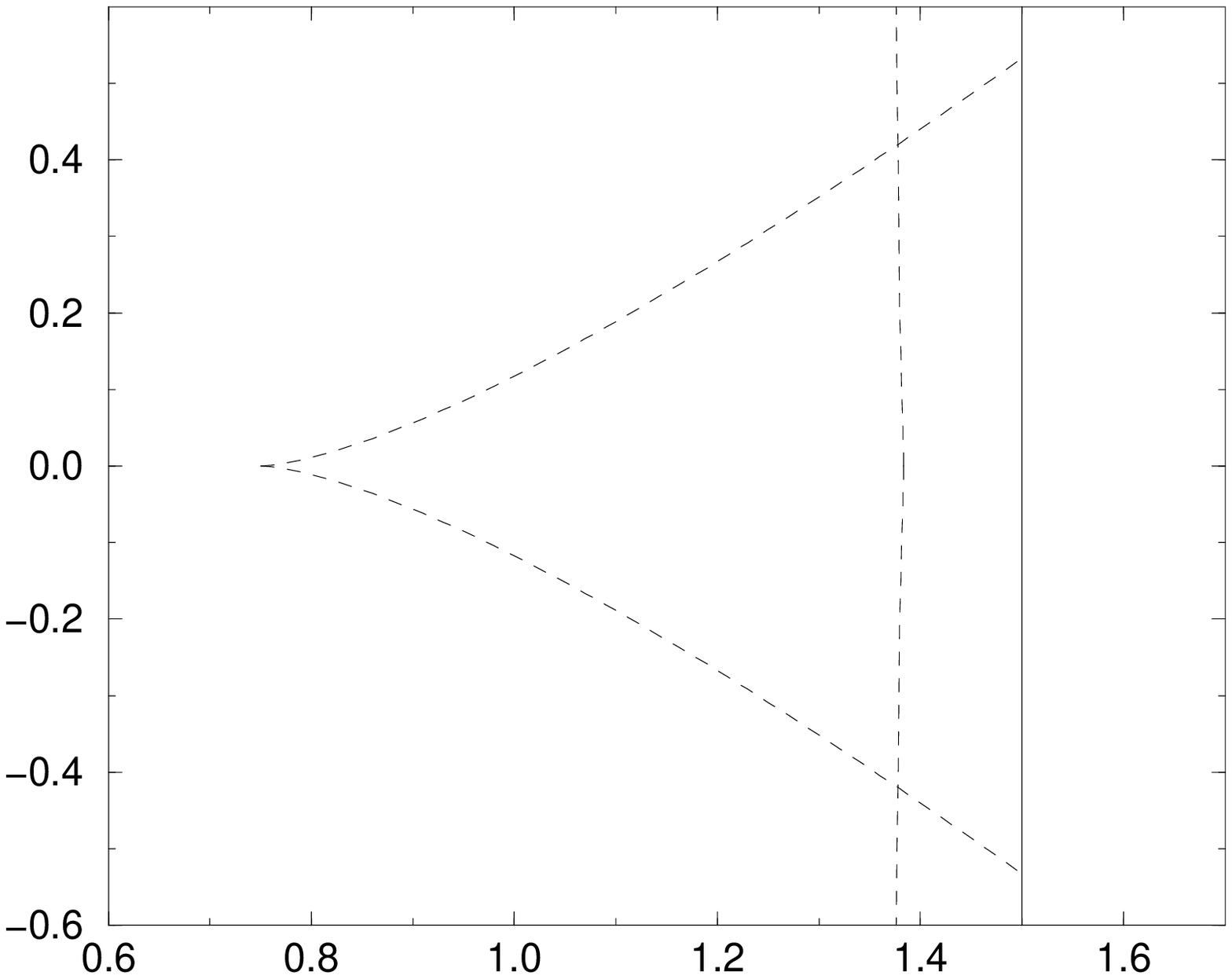}}
\put(140,  50){\large $\tilde{\beta}\overline{\mu}$}
\put(209,-9){\large $\beta J$}
 \put(180, 75){$S1$}  \put(180, 25){$S1$}
 \put(210, 50){$S2$}
 \put(205,  87){$S1\!+\!C1$}\put(225,89){\line(1,0){15}}
   \put(252, 50){$CO$}
  \put(205,13){$S1\!+\!C1$}\put(225,15){\line(1,0){15}}
   \put(239,56){$S2$}\put(239,50){$+$}\put(239,44){$C2$}
\end{picture}
\vspace*{7mm}
 \caption{Phase diagrams for $q=3$ and
$P(\mu)=\delta[\mu-\overline{\mu}]$. The possible phases are
$\{S\ell\}$ (swollen phases with $\ell$ possible locally stable
values of $p$), $\{C\ell\}$ (compact phases with $\ell$ possible
locally stable values of $p$), $\{S\ell\plus C\ell\}$ (phases with
$2\ell$ possible locally stable states of $p$, half of which are
compact), and $CO$ (compact states only). All transitions are
first order, except for the one at $\beta J=\frac{3}{2}$ (which is
second order). Upper left diagram: $n=\frac{1}{2}$. Upper right
diagram: $n=1$. Lower left diagram: $n=\frac{6}{5}$. Lower right
diagram: $n=2$. } \label{fig:singlemupdiagram1}
\end{figure}

 Since for both $T\to
0$ and $T\to\infty$ there are only two non-identical values for
the order parameters $L_\phi$, we expect the solution to be of the
form $\bL=(L_1,L_2,L_2)$ at all temperatures. We thus insert the
ansatz $\{L_{\pm
2\pi/3}=\frac{1}{3}(2p+Z),L_0=\frac{1}{3}(2p-2Z)\}$ into our
saddle-point equations and find
\be
p= \frac{e^{-\beta
n[\overline{\mu}-\frac{2}{3}Jp]}(2e^{\frac{1}{3}\beta J
Z}+e^{-\frac{2}{3}\beta J Z})^n- e^{\beta
n[\overline{\mu}-\frac{2}{3}Jp]}(2e^{-\frac{1}{3}\beta J
Z}+e^{\frac{2}{3}\beta J Z})^n }{e^{-\beta
n[\overline{\mu}-\frac{2}{3}Jp]}(2e^{\frac{1}{3}\beta J
Z}+e^{-\frac{2}{3}\beta J Z})^{n}+ e^{\beta
n[\overline{\mu}-\frac{2}{3}Jp] } (2e^{-\frac{1}{3}\beta J
Z}+e^{\frac{2}{3}\beta J Z})^n} \label{eq:q=3_peq} \ee
\be
Z=F[Z;p]~~~~~~~~~~F[Z;p]=\frac{2p[1-\cosh(\beta JZ)]+6\sinh(\beta
JZ)}{5+4\cosh(\beta JZ)} \label{eq:ZFZ} \ee with a corresponding
simplification of the bifurcation condition
(\ref{eq:q=3transitions}). For $Z=0$ these equations bring us back
to the swollen state $L_\phi=\frac{2}{3}p$, with $p=\tanh[\beta
n(\frac{2}{3}Jp-\overline{\mu})]$. Comparison with the
corresponding equation for $q=2$ shows that the properties of the
 swollen state with $q=3$ can  be obtained from those derived for
$q=2$ via the substitution $J\to \frac{2}{3}J$. This gives
\begin{eqnarray*}
{\rm swollen~state:}~~~~~~ \tilde{\beta}J
>\frac{3}{2}~~{\rm and}~~|\overline{\mu}|<\mu_c: &~~~& {\rm two~local~minima~}p\\
\hspace*{34mm}{\rm elsewhere:} &~~&  {\rm one~local~minimum~}
\end{eqnarray*}
with
\be
\tilde{\beta}\mu_c=\sqrt{\frac{2}{3}\tilde{\beta}J}\sqrt{\frac{2}{3}\tilde{\beta}J
\minus 1} -\frac{1}{2} \log\left[
\frac{\sqrt{\frac{2}{3}\tilde{\beta}J}\plus
\sqrt{\frac{2}{3}\tilde{\beta}J \minus
1}}{\sqrt{\frac{2}{3}\tilde{\beta}J}\minus
\sqrt{\frac{2}{3}\tilde{\beta}J\minus 1}} \right]
 \label{eq:phaseequation_q3}
\ee The transitions at $\overline{\mu}=\pm \mu_c$ are first-order,
except for the point
$(\tilde{\beta}J,\tilde{\beta}\overline{\mu})=(\frac{3}{2},0)$
where they are second-order. Since the swollen state is locally
unstable (against collapsed solutions) for $\beta J>\frac{3}{2}$,
the transitions at $\overline{\mu}=\pm \mu_c$ can be observed only
for $n>1$. It follows from $F[Z;p]=\frac{2}{3}\beta J
Z+\order(Z^2)$ that the swollen state, $Z=0$, indeed destabilizes
at $\beta J=\frac{3}{2}$ in favor of a collapsed state of the type
above, with $Z\neq 0$.

The main qualitative change  observed for $q=3$ is the emergence
of prominent transitions to compact states before the temperature
where the swollen state becomes locally unstable. These are in
fact also found to be present in the long range limit of
\cite{Skantzos}, where there is no genetic dynamics and where $p$
is a control parameter, but far less prominently and with very
small attraction domains of the associated newly created free
energy minima. Note that for $n\to 0$ our present model will not
exhibit these first order transitions, since this would reduce our
system to the $p=0$ case of \cite{Skantzos} (the first order
transitions  are found only for $p\neq 0$).

\section{Results for zero-average binary genetic forces}

Our second example is the symmetric binary distribution
$P(\mu)=\frac{1}{2}\delta[\mu\plus\sigma]+\frac{1}{2}\delta[\mu\minus\sigma]$,
with $\sigma\geq 0$.  Here
 the functionality potential
$V(\bfeta)$ in the genetic dynamics discourages the biologically
undesirable single-species states with $p=\pm 1$. Again we first
inspect the limits $T\to 0$ and $n\to\infty$.

\subsection{The deterministic limits: $T\to 0$ and $n\to\infty$}

We first calculate the ground state (deterministic evolution of
configurational angles: $T=0$).  For the present choice of
$P(\mu)$ one finds for
 the function $\epsilon(p)$
in (\ref{eq:groundstate_nonsym}):
\bd
  \epsilon(p)=
\frac{1}{2} p^2 -\frac{1}{2} - \frac{1}{2}|p\!- \frac{\sigma}{J}|
- \frac{1}{2}|p\!+ \frac{\sigma}{J}| \ed Upon working out the
details in the three regimes
$\{p<-\sigma/J,~|p|<\sigma/J,~p>\sigma/J\}$ one finds that for
$\sigma>0$ there is always a local minimum of $\epsilon(p)$ at
$p=0$, with $E=-\frac{1}{2}J-\sigma$, and that for $\sigma<J$
there are two additional local minima of $\epsilon(p)$ at $p=\pm
1$ with $E=-J$. The latter will replace $p=0$ as the global
minimum for $\sigma<\frac{1}{2}J$:
\be
\begin{array}{lllllll}  \sigma>J/2: &~~~&
E_0=-\frac{1}{2}J-\sigma,&& p=0, && \bL=(-1,0,\ldots,0,1)\\[1mm]
\sigma<J/2: &~~~& E_0=-J, && p=\pm 1,&& \bL=2p(1,0,\ldots,0)
\end{array}
\label{eq:groundstate_binary}
 \ee
For $\sigma>\frac{1}{2}J$ the ground state
(\ref{eq:groundstate_binary}) is one where half of the monomers
are hydrophilic and half are hydrophobic (in line with the
requirements of the  potential $V(\bfeta)$), and where the two
species are perfectly separated with all hydrophobic monomers
clustered at one site, and all hydrophilic ones clustered at
another. The origin of the (undesirable) single-species minima at
$p=\pm 1$, for $\sigma<J/2$,
 can be understood upon realizing that for $\sigma\to 0$
one returns to the $\delta$-distribution studied earlier, with
$\overline{\mu}=0$. For $\sigma=J/2$ both types of states
(multiple-species and single-species) give energetically
equivalent minima.

 For $n\to\infty$ (deterministic genetic dynamics) we get
\begin{eqnarray}
 \lim_{n\to\infty} f_{\rm RS}& =&
 {\rm min}_{\{\ell_\phi\}}\left\{
 \frac{J}{4}\sum_{\phi}\ell^2_\phi
 -
 \frac{1}{2\beta}\log(\sum_\phi e^{\beta J
\ell_\phi}) -
 \frac{1}{2\beta}\log(\sum_\phi e^{-\beta J \ell_\phi}) \right.
\nonumber\\ &&\left.
 -\frac{1}{2}
\left|\sigma\minus \frac{1}{2\beta}\log\left[\frac{\sum_\phi
e^{\beta J \ell_\phi}}{\sum_\phi e^{-\beta J
\ell_\phi}}\right]\right|
 -\frac{1}{2}
\left|\sigma\plus \frac{1}{2\beta}\log\left[\frac{\sum_\phi
e^{\beta J \ell_\phi}}{\sum_\phi e^{-\beta J
\ell_\phi}}\right]\right|
 \right\}
\end{eqnarray}
\begin{eqnarray}
p&=& -\frac{1}{2}\sgn\left[\sigma\minus
\frac{1}{2\beta}\log\left[\frac{\sum_\phi e^{\beta J
L_\phi}}{\sum_\phi e^{-\beta J L_\phi}}\right]\right]
+\frac{1}{2}\sgn\left[\sigma\plus
\frac{1}{2\beta}\log\left[\frac{\sum_\phi e^{\beta J
L_\phi}}{\sum_\phi e^{-\beta J L_\phi}}\right]\right]
\end{eqnarray}
We now have five candidate solutions
$p\in\{-1,-\frac{1}{2},0,\frac{1}{2},1\}$ (due to the symmetry
under $\{L_\phi,p\}\to \{\minus L_\phi,\minus p\}$ of the
minimization problem, the $p=\pm 1$ and $p=\pm \frac{1}{2}$
solutions are mutually equivalent). The potentially stable ones
are $p\in\{-1,0,1\}$, separated by the unstable fixed-points
$p=\pm \frac{1}{2}$:
\begin{eqnarray*}
p=1:~~~&&  L_\psi= \frac{ 2 e^{\beta J L_\psi}}{\sum_\phi e^{\beta
JL_\phi}},~~~~~~~~~~~~~~~~~~~~~~~~~\frac{1}{2\beta}\log\left[\frac{\sum_\phi
e^{\beta J L_\phi}}{\sum_\phi e^{-\beta J L_\phi}}\right]
>\sigma
\nonumber
\\
&& \lim_{n\to\infty} f_{\rm RS} =
 \frac{J}{4}\sum_{\phi}L^2_\phi
 -
 \frac{1}{\beta}\log(\sum_\phi e^{\beta J
L_\phi})
 \\
 p=0:~~~&&
 L_{\psi}= \frac{e^{\beta J L_\psi}}{\sum_\phi e^{\beta J L_\phi}}
 -\frac{e^{-\beta J L_\psi}}{
\sum_\phi e^{-\beta J L_\phi}},~~~~~~
\left|\frac{1}{2\beta}\log\left[\frac{\sum_\phi e^{\beta J
L_\phi}}{\sum_\phi e^{-\beta J L_\phi}}\right]\right|<\sigma
\nonumber
\\
&& \lim_{n\to\infty} f_{\rm RS} = \frac{J}{4}\sum_{\phi}L^2_\phi
 -
 \frac{1}{2\beta}\log(\sum_\phi e^{\beta J
L_\phi}) -
 \frac{1}{2\beta}\log(\sum_\phi e^{-\beta J
L_\phi})-\sigma
\\
 p=-1:~~~&& L_\psi= \frac{ -2 e^{-\beta J
L_\psi}}{\sum_\phi e^{-\beta
JL_\phi}},~~~~~~~~~~~~~~~~~~~~~~~~\frac{1}{2\beta}\log\left[\frac{\sum_\phi
e^{\beta J L_\phi}}{\sum_\phi e^{-\beta J L_\phi}}\right]<-\sigma
\nonumber
\\ &&
 \lim_{n\to\infty} f_{\rm RS} =
 \frac{J}{4}\sum_{\phi}L^2_\phi -
 \frac{1}{\beta}\log(\sum_\phi e^{-\beta J L_\phi})
 \end{eqnarray*}
 For the swollen state $L_\phi=2p/q$ $\forall \phi$
this implies: for $\sigma>2J/q$ there is only the solution $p=0$,
for $\sigma<2J/q$ also the solutions $p=\pm 1$ are  locally stable
(but with a lower free energy than the $p=0$ state only when
$\sigma<J/q$).

The general picture emerging from the above two limit cases is, as
expected, that for sufficiently large $\sigma$ the single-species
states $p=\pm 1$ (which continue to be energetically favorable for
the fast process) will indeed be replaced by a biologically more
interesting state with equal fractions of hydrophilic  and
hydrophobic monomers.

\subsection{Phase diagrams for $q=2$}

\begin{figure}[t]
\vspace*{18mm}\hspace*{1mm} \setlength{\unitlength}{0.62mm}
\begin{picture}(0,70)
\put(0,-10){\epsfxsize=140\unitlength\epsfbox{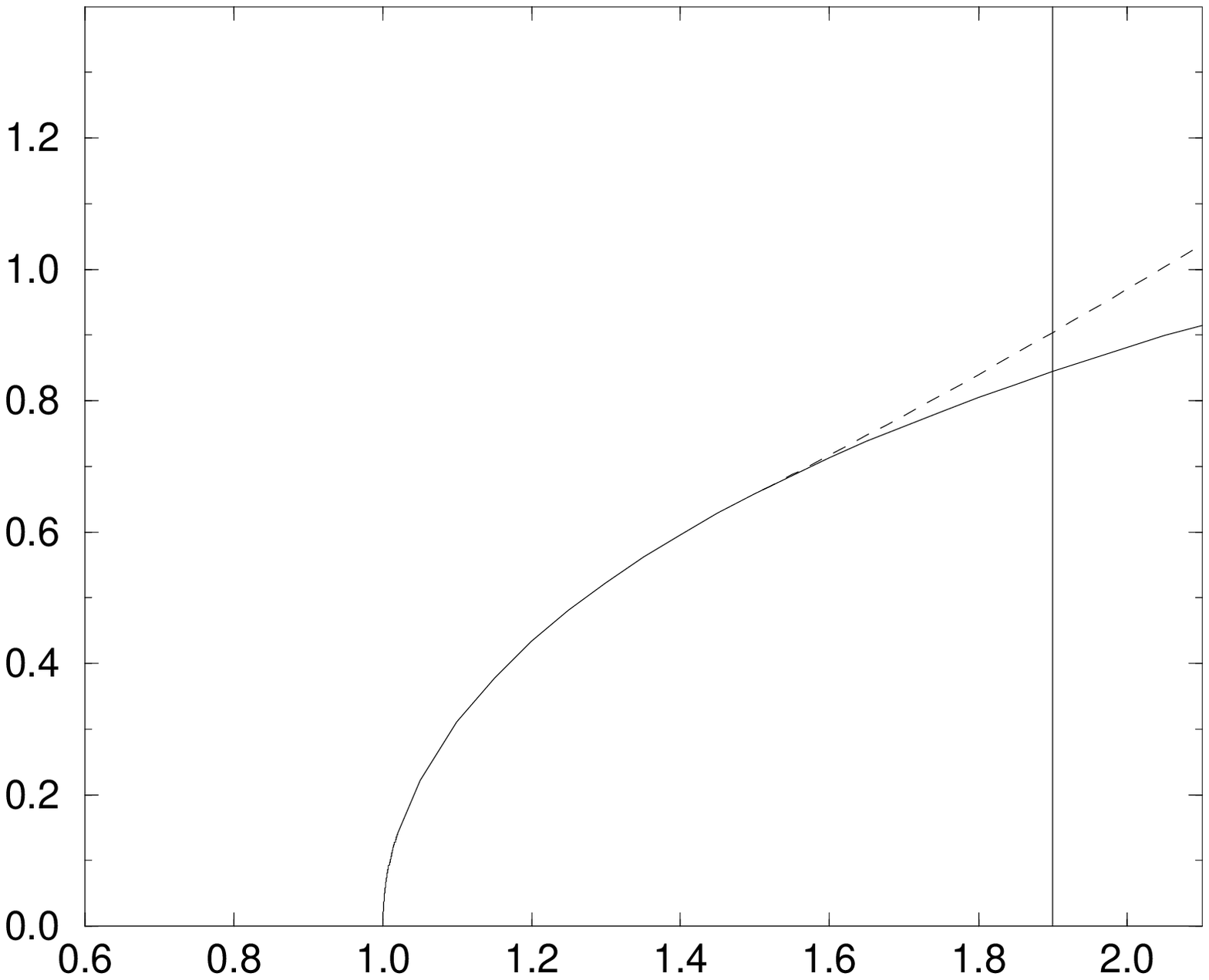}}
\put(0, 50){\large $\tilde{\beta}\sigma$} \put(69,-9){\large
$\tilde{\beta} J$} \put(45,60){$S1$}  \put(116,81){$C1$}
\put(80,25){$S2$} \put(116,35){$C2$}
 \put(95,71){$C3$}
 \put(103,72){\line(4,-1){18}}
 \put(79,65){$S3$}
 \put(87,66){\line(4,-1){22}}
\put(140,-10){\epsfxsize=140\unitlength\epsfbox{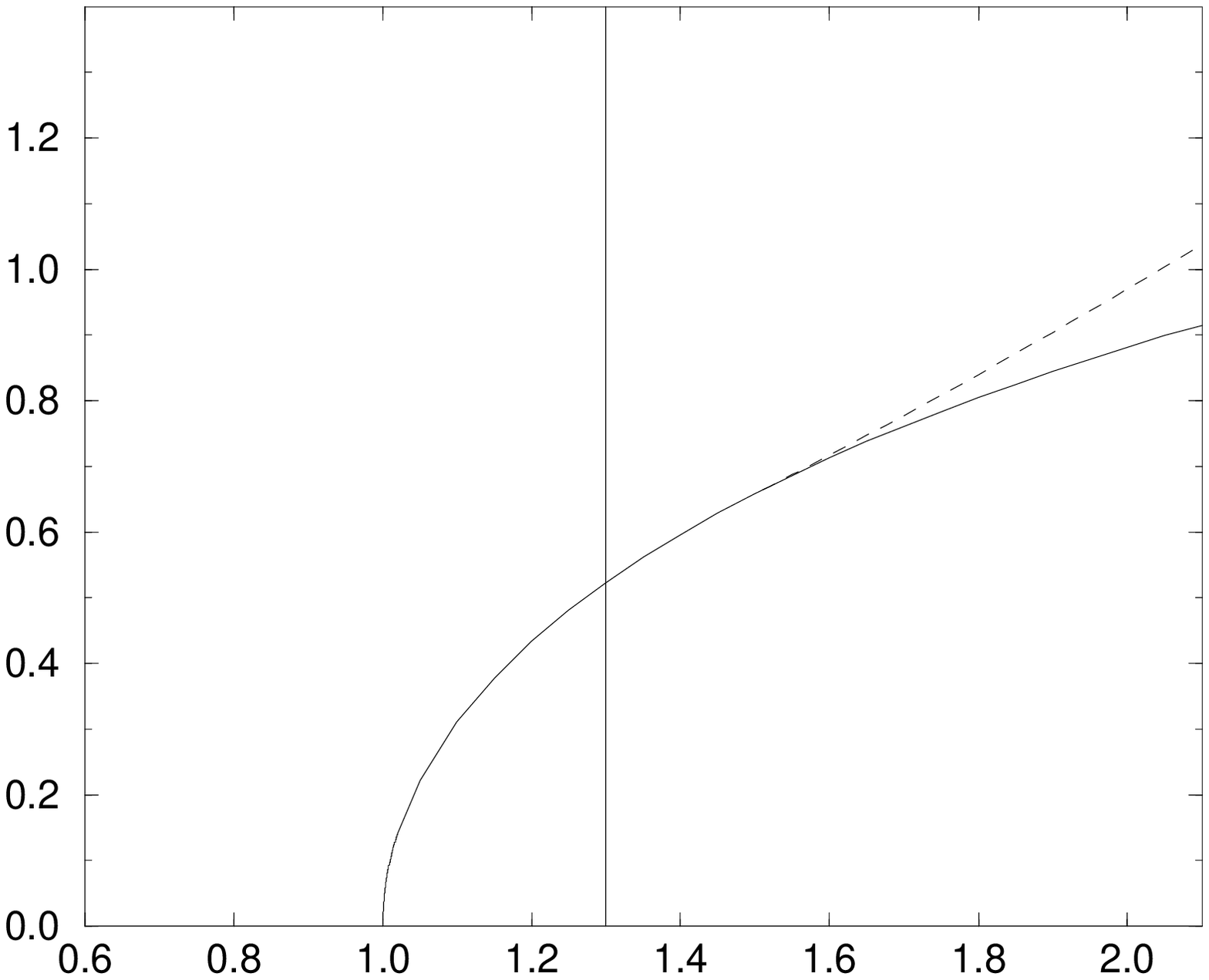}}
 \put(140,  50){\large $\tilde{\beta}\sigma$}
\put(209,-9){\large $\tilde{\beta}J$} \put(185,60){$S1$}
\put(219,80){$C1$} \put(199,20){$S2$} \put(236,30){$C2$}
 \put(235,70){$C3$}
 \put(243,71){\line(4,-1){18}}
\end{picture}
\vspace*{10mm} \caption{Phase diagrams for $q=2$ and
$P(\mu)=\frac{1}{2}\delta[\mu\minus
\sigma]+\frac{1}{2}\delta[\mu\plus\sigma]$. The six possible
phases are $\{S1,S2,S3\}$ (swollen phases with $1,~2$ or $3$
possible locally stable values of $p$)  and $\{C1,C2,C3\}$
(compact phases with $1,~2$ or $3$ possible locally stable values
of $p$). The $S\ell\to C\ell$  transitions (at $\tilde{\beta}J=n$)
as well as the $S1\to S2$ and $C1\to C2$ transitions are all
second order. The $S1\to S3$ and $C1\to C3$ transitions are first
order. Left diagram: $n=5$. Middle diagram: $n=\frac{5}{4}$. Right
diagram: $n=\frac{1}{2}$. Note that $\tilde{\beta}=\beta n$, that
 phase $S2$ exists only for $n>1$, and that phase $S3$ exists only for
$n>\frac{3}{2}$.} \label{fig:binaryphasediagramq=2}
\end{figure}

Now the order parameter equation becomes
\be
 p=G(p),~~~~~~~~G(p)=\frac{1}{2}\tanh[\tilde{\beta}(Jp\minus \sigma)]+\frac{1}{2}\tanh[\tilde{\beta}(Jp\plus\sigma)]
\label{eq:peqn_binary} \ee Due to the symmetry $P(\mu)=P(-\mu)$ we
always find the state $p=0$ (equal fractions of hydrophobic and
hydrophilic monomers). We expand the function $G(p)$ in powers of
$p$, \bd
G(p)=\tilde{\beta}J[1-\tanh^2(\tilde{\beta}\sigma)].p-\frac{1}{3}(\tilde{\beta}J)^3\left[
1-4\tanh^2(\tilde{\beta}\sigma)+3\tanh^4(\tilde{\beta}\sigma)\right].p^3+\order(p^5)
\ed and conclude that the state $p=0$ de-stabilizes when
$\tanh^2(\tilde{\beta}\sigma)=1-1/\tilde{\beta}J$, or,
equivalently, at
\be
\tilde{\beta}\sigma_c=\frac{1}{2}\log\left[
\frac{\sqrt{\tilde{\beta}J}+\sqrt{\tilde{\beta}J\minus 1}}
{\sqrt{\tilde{\beta}J}-\sqrt{\tilde{\beta}J\minus 1}} \right]
\label{eq:2nd_order_binary} \ee and that this corresponds to a
true second-order transition as long as $\tilde{\beta}J<3/2$ (so
that $ G^{\prime\prime\prime}(0)$
 is negative).
We note that \begin{eqnarray*}
\tilde{\beta}\sigma_c=(\tilde{\beta}J\minus 1)^{1/2}
+\order(\tilde{\beta}J\minus 1) &~~~~~& (\tilde{\beta}J\to 1)\\
\tilde{\beta}\sigma_c=\frac{1}{2}\log(\tilde{\beta}J)+\order(1) &
~~~~~& (\tilde{\beta}J\to \infty)
\end{eqnarray*}
 For $\tilde{\beta}J>3/2$, however, there
are first-order transitions away from the state $p=0$ at
temperatures higher than the one defined by
(\ref{eq:2nd_order_binary}). The locations in the phase diagram of
the  latter, which start at the triple point
$(\tilde{\beta}J,\tilde{\beta}\sigma)=(\frac{3}{2},{\rm
arctanh}[1/\sqrt{3}])$, are to be solved (numerically) from the
coupled transcendental equations $p=G(p)$ and $1=G^\prime(p)$,
i.e.
\begin{eqnarray}
p&=&\frac{1}{2}\tanh[\tilde{\beta}(Jp\minus
\sigma)]+\frac{1}{2}\tanh[\tilde{\beta}(Jp\plus\sigma)]\\
1&=&\tilde{\beta}J\left\{1-\frac{1}{2}\tanh^2[\tilde{\beta}(Jp\minus\sigma)]
-\frac{1}{2}\tanh^2[\tilde{\beta}(Jp\plus\sigma)]\right\}
\label{eq:peqnderivative_binary}
\end{eqnarray}
It is now (in contrast to the case of $\delta$-distributed genetic
forces) possible to have {\em three} locally stable values for the
average polarity $p$ (one of which is zero, the other two are
different in sign only). This exhausts the phases and transitions:
there is one second order transition at $\beta J=1$ from a swollen
state with uniformly distributed monomer orientations to a compact
state with separation of polarity types,
 two first-order transition lines (in
the region $\beta n J<\frac{3}{2}$) marking the creation of
multiple locally stable values for the average polarity together
with de-stabilization of the state $p=0$, and two first order
transition lines where two nonzero values of $p$ bifurcate
discontinuously (without affecting the stability of $p=0$). These
lines are shown in the $(\tilde{\beta}J,\tilde{\beta}\sigma)$
phase diagram, in figure \ref{fig:binaryphasediagramq=2}. We
denote swollen phases with $\ell$ locally stable values of $p$ as
$S\ell$, and
 compact phases with $\ell$  locally stable values of
$p$ as $C\ell$. The number of possible phases is again found to
depend explicitly on $n$: phase $S2$ exists only for $n>1$, and
$S3$ exists only for $n>\frac{3}{2}$.

\subsection{Phase diagrams for $q=3$}

\begin{figure}[t]
\vspace*{18mm}\hspace*{1mm} \setlength{\unitlength}{0.62mm}
\begin{picture}(0,70)
\put(0,-10){\epsfxsize=140\unitlength\epsfbox{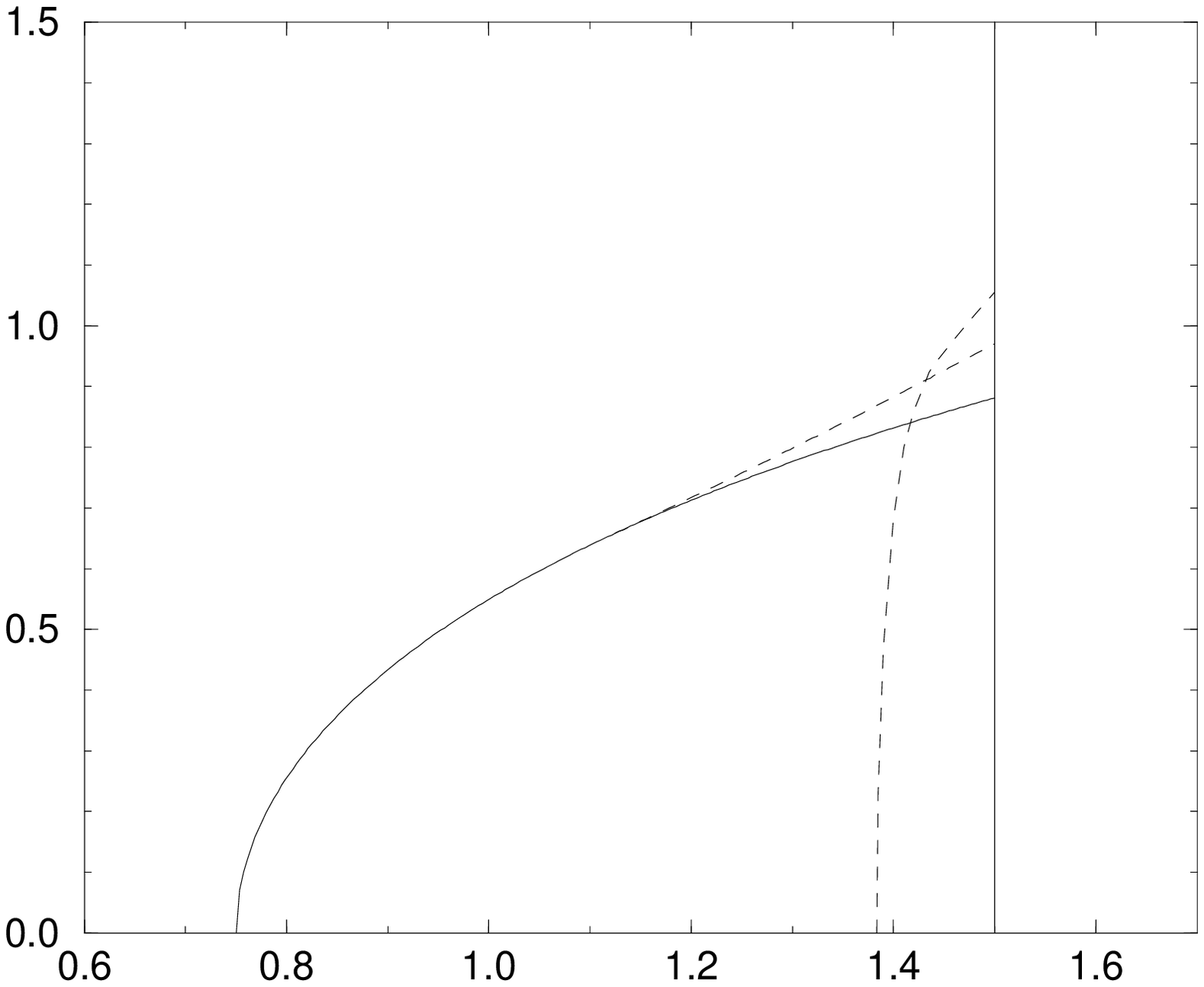}}
\put( 0,  48){\large $\tilde{\beta}\sigma$} \put( 70,  -9){\large
$\beta{J}$} \put( 40,  70){$S1$} \put( 65, 25){$S2$} \put( 113,
50){$CO$}
 \put(99,36){$S2$}\put(99,30){$+$}\put(99,24){$C2$}
 \put(140,-10){\epsfxsize=140\unitlength\epsfbox{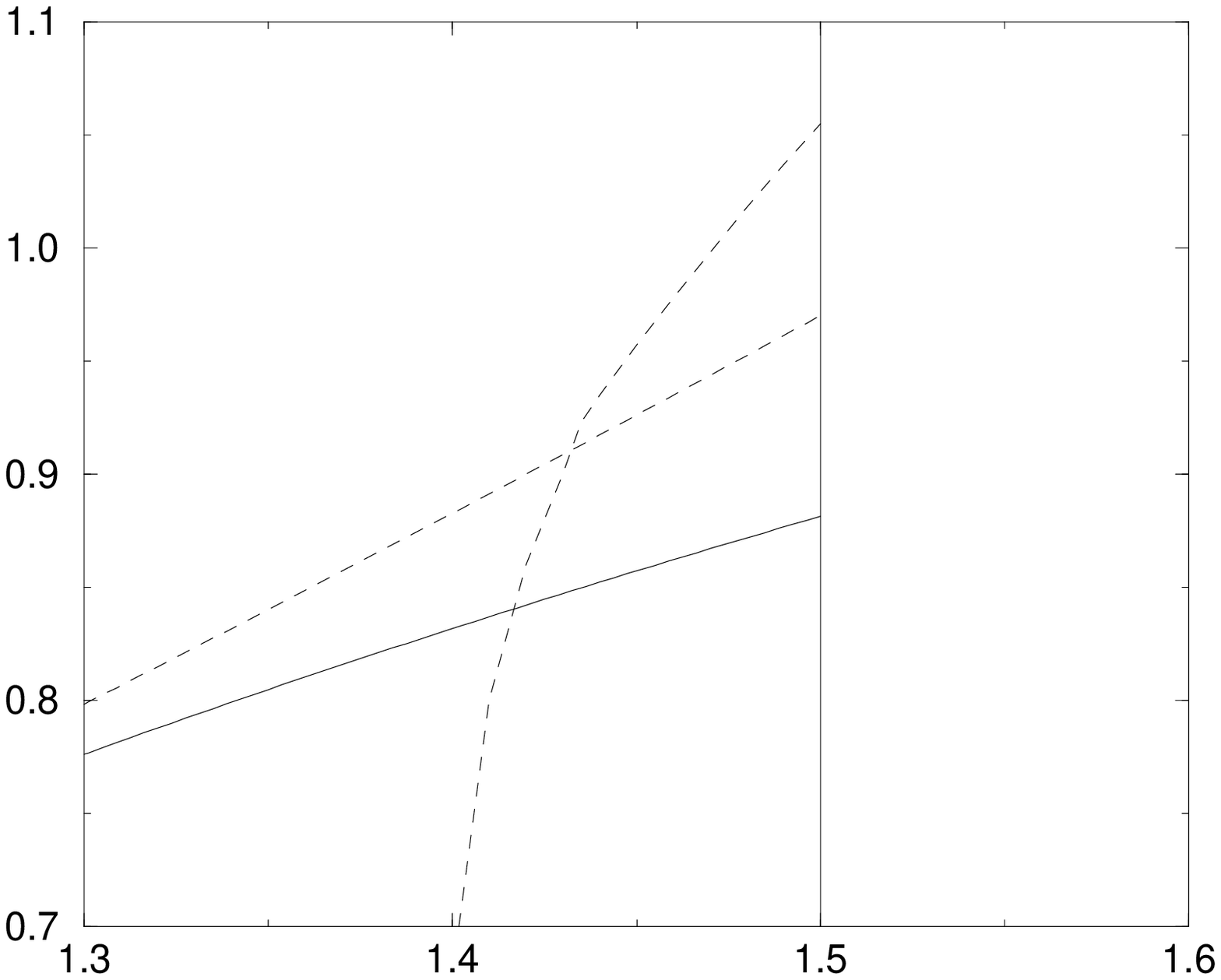}}
\put(140,  48){\large $\tilde{\beta}\sigma$} \put(210,  -9){\large
$\beta{J}$} \put(170,  80){$S1$} \put(170,  15){$S2$} \put(194,
40){$S3$}  \put(211, 49){$S3\plus C3$}  \put(245, 50){$CO$}
 \put(207,85){$S1\plus C1$}\put(216,83){\line(1,-1){10}}
 \put(205,15){$S2\plus C2$}
\end{picture}
\vspace*{8mm} \caption{ Phase diagrams for $q=3$ and
$P(\mu)=\frac{1}{2}\delta[\mu\minus
\sigma]+\frac{1}{2}\delta[\mu\plus\sigma]$, with $n=2$. The
possible phases are $\{S\ell\}$ (swollen phases with $\ell$
possible locally stable values of $p$), $\{C\ell\}$ (compact
phases with $\ell$ possible locally stable values of $p$),
$\{S\ell\plus C\ell\}$ (phases with $2\ell$ possible locally
stable states of $p$, half of which are compact), and $CO$
(compact states only). First order transitions are indicated by
dashed curves, second order ones by solid curves. The right
diagram is an enlargement of one of the two areas in the left
diagram where all lines nearly meet. } \label{fig:binaryq3}
\end{figure}

For the force distribution $P(\mu)=\frac{1}{2}\delta[\mu\minus
\sigma]+\frac{1}{2}\delta[\mu\plus\sigma]$ we know the ground
state to be $\bL=(-1,0,1)$ for   $\sigma>J/2$ (with $p=0$) and
$\bL=2p(1,0,0)$ for $\sigma<J/2$ (with $p=\pm 1$). We therefore
expect the solution to be of the form $\bL=(L_1,L_2,L_2)$ at all
temperatures {\em at most} when $\sigma<J/2$. Inserting $\{L_{\pm
2\pi/3}=\frac{1}{3}(2p+Z),L_0=\frac{1}{3}(2p-2Z)\}$ into our
saddle-point equations now leads to
\begin{eqnarray}
p&=& \frac{1}{2}\frac{e^{-\beta
n[\sigma-\frac{2}{3}Jp]}(2e^{\frac{1}{3}\beta J
Z}+e^{-\frac{2}{3}\beta J Z})^n- e^{\beta
n[\sigma-\frac{2}{3}Jp]}(2e^{-\frac{1}{3}\beta J
Z}+e^{\frac{2}{3}\beta J Z})^n }{e^{-\beta
n[\sigma-\frac{2}{3}Jp]}(2e^{\frac{1}{3}\beta J
Z}+e^{-\frac{2}{3}\beta J Z})^{n}+ e^{\beta
n[\sigma-\frac{2}{3}Jp] } (2e^{-\frac{1}{3}\beta J
Z}+e^{\frac{2}{3}\beta J Z})^n}\nonumber
\\
&&+\frac{1}{2} \frac{e^{\beta
n[\sigma+\frac{2}{3}Jp]}(2e^{\frac{1}{3}\beta J
Z}+e^{-\frac{2}{3}\beta J Z})^n- e^{-\beta
n[\sigma+\frac{2}{3}Jp]}(2e^{-\frac{1}{3}\beta J
Z}+e^{\frac{2}{3}\beta J Z})^n }{e^{\beta
n[\sigma+\frac{2}{3}Jp]}(2e^{\frac{1}{3}\beta J
Z}+e^{-\frac{2}{3}\beta J Z})^{n}+ e^{-\beta
n[\sigma+\frac{2}{3}Jp] } (2e^{-\frac{1}{3}\beta J
Z}+e^{\frac{2}{3}\beta J Z})^n}
 \end{eqnarray}
\be
Z=F[Z;p]~~~~~~~~~~F[Z;p]=\frac{2p[1-\cosh(\beta JZ)]+6\sinh(\beta
JZ)}{5+4\cosh(\beta JZ)} \ee (with a corresponding simplification
of the bifurcation condition (\ref{eq:q=3transitions})). For $Z=0$
these equations bring us back to the swollen state
$L_\phi=\frac{2}{3}p$, but now with \bd p=\frac{1}{2}\tanh[\beta
n(\frac{2}{3}Jp-\sigma)]+\frac{1}{2} \tanh[\beta
n(\frac{2}{3}Jp+\sigma)]
 \ed Comparison with the
corresponding equation for $q=2$ shows that, for
$\sigma<\frac{1}{2}J$, the properties of the
 swollen state with $q=3$ can again be obtained from those derived for
$q=2$ via the substitution $J\to \frac{2}{3}J$. This tells us that
the state $p=0$ (which always solves the saddle-point equation)
de-stabilizes at
\be
\tilde{\beta}\sigma_c=\frac{1}{2}\log\left[
\frac{\sqrt{\frac{2}{3}\tilde{\beta}J}+\sqrt{\frac{2}{3}\tilde{\beta}J\minus
1}}
{\sqrt{\frac{2}{3}\tilde{\beta}J}-\sqrt{\frac{2}{3}\tilde{\beta}J\minus
1}} \right] \label{eq:2nd_order_binary_q3} \ee and that this
corresponds to a true second-order transition as long as
$\tilde{\beta}J<9/4$.
 For $\tilde{\beta}J>9/4$, however, there
are first-order transitions away from the state $p=0$ at
temperatures higher than the one defined by
(\ref{eq:2nd_order_binary_q3}), which start at the triple point
$(\tilde{\beta}J,\tilde{\beta}\sigma)=(\frac{9}{4},{\rm
arctanh}[1/\sqrt{3}])$, to be solved (numerically) from the
coupled equations
\begin{eqnarray}
p&=&\frac{1}{2}\tanh[\tilde{\beta}(\frac{2}{3}Jp\minus
\sigma)]+\frac{1}{2}\tanh[\tilde{\beta}(\frac{2}{3}Jp\plus\sigma)]\\
1&=&\frac{2}{3}\tilde{\beta}J\left\{1-\frac{1}{2}\tanh^2[\tilde{\beta}(\frac{2}{3}Jp\minus\sigma)]
-\frac{1}{2}\tanh^2[\tilde{\beta}(\frac{2}{3}Jp\plus\sigma)]\right\}
\label{eq:peqnderivative_binary_q3}
\end{eqnarray}
Thus we can have, even for the swollen state, three locally stable
values for the average polarity $p$ (one of which is zero, the
other two are different in sign only).
 Since the swollen state is locally unstable (against
collapsed solutions) for $\beta J>\frac{3}{2}$, however, the above
transitions can be seen  only for $n>1$. Since
$F[Z;p]=\frac{2}{3}\beta J Z+\order(Z^2)$, the swollen state,
$Z=0$ destabilizes at $\beta J=\frac{3}{2}$ in favor of a
collapsed state of the type $\bL=(L_1,L_2,L_2)$.

We show the resulting phase diagram (complemented by numerical
analysis of the transitions) in figure \ref{fig:binaryq3}, for
$n=2$. As a result of the symmetry $P(\mu)=P(\minus \mu)$,  here
the $S1$ phase always has $p=0$ (in contrast to e.g. the case of
$\delta$-distributed forces), whereas $C1$ has $p\neq 0$.
 The $S2$ and $C2$ phases also involve $p\neq 0$ (with the two possible values different in sign only).
 The only phase that exists for large values of $\sigma$ is
$S1$. One notes again the crucial dependence on $n$ of the
existence of some of the phases (clearly evident in the equations
above). It is now clear that upon choosing a sufficiently large
value for $\sigma$ one can eliminate the biologically undesirable
single species states at any temperature $T$.

\section{Results for Gaussian-distributed genetic forces}

\begin{figure}[t]
\vspace*{0mm} \setlength{\unitlength}{0.9mm} \hspace*{40mm}
\begin{picture}(100,85)
 \put(7,20){\epsfxsize=91\unitlength\epsfbox{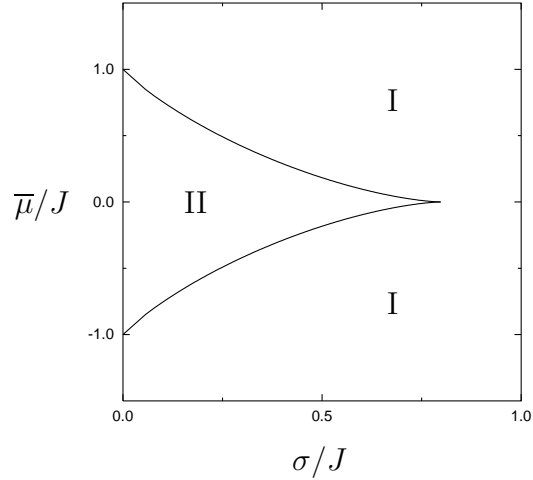}}
 \put(-1,50){\large $\overline{\mu}/J$}   \put(40,12){\large $\sigma/J$}
 \put(54,65){\large I}
 \put(54,35){\large I}
 \put(24,50){\large II}
\end{picture}
\vspace*{-11mm} \caption{Onset of multiplicity of locally stable
states at $T=0$, for Gaussian-distributed forces $\mu_i$ (with
average $\overline{\mu}$ and variance $\sigma^2$). Solid curves:
critical lines $\pm \overline{\mu}_c/J$ in parameter space as
defined in (\ref{eq:groundstate_lines}), terminating at $\sigma/J=
\sqrt{2/\pi}$. Region I: a single local and global minimum (the
ground state), with $\sgn[p]=-\sgn[\overline{\mu}]$. Region II: an
additional local minimum (with energy higher than the ground
state), with $\sgn[p]=\sgn[\overline{\mu}]$.}
\label{fig:groundstate}
\end{figure}

 Our third example is the Gaussian
distribution
$P(\mu)=[2\pi\sigma^2]^{-\frac{1}{2}}e^{-\frac{1}{2}[\mu-\overline{\mu}]^2/\sigma^2}$.
Again, provided $\sigma$ is sufficiently large, the genetic
functionality potential discourages the biologically undesirable
single-species states. For $\sigma\to \infty$ it follows
immediately from  (\ref{eq:RSp}) that $p=0$.

\subsection{The deterministic limits: $T\to 0$ and $n\to\infty$}

Working out the relevant Gaussian integral in
(\ref{eq:groundstate_nonsym}) now gives us \be
  \epsilon(p)=
\frac{1}{2} p^2 -\frac{1}{2} -
\frac{\sigma}{J}F\left[\frac{pJ-\overline{\mu}}{\sigma}\right]
~~~~~~~~~~ F[x]=x ~{\rm
Erf}[\frac{x}{\sqrt{2}}]+\sqrt{\frac{2}{\pi}}~ e^{-\frac{1}{2}x^2}
\label{eq:defineF} \ee It follows that $d \epsilon(p)/dp=p-{\rm
Erf}[ (pJ\minus \overline{\mu})/\sigma\sqrt{2}]$. Graphical
inspection of the functions $\epsilon(p)$ and $d\epsilon(p)/dp$
reveals that for $\overline{\mu}\neq 0$ the global minimum of
$\epsilon(p)$ (i.e. the true ground state) is always at a value of
$p$ with $\sgn[p]=-\sgn[\overline{\mu}]$, but that  for
sufficiently small $\sigma$ there will be an additional local
minimum (bifurcating discontinuously, together with a
corresponding local maximum) with $\sgn[p]=\sgn[\overline{\mu}]$,
provided $|\overline{\mu}|<J$. The bifurcation point is calculated
by solving simultaneously the equations $d \epsilon(p)/d p=d^2
\epsilon(p)/d p^2=0$. Working out these equations and eliminating
$p$ leads us to the two critical lines $\overline{\mu}=\pm
\overline{\mu}_c(\sigma)$, which separate a region with a single
local minimum from a region with two local minima:
\be
\frac{\overline{\mu}_c}{J}= {\rm
Erf}\left[\frac{1}{\sqrt{2}}\sqrt{\log\left[\frac{2J^2}{\pi\sigma^2}\right]}\right]
-
\frac{\sigma}{J}\sqrt{\log\left[\frac{2J^2}{\pi\sigma^2}\right]}
\label{eq:groundstate_lines}
 \ee The result is shown in figure
\ref{fig:groundstate}. For $\sigma\to 0$ one finds
$\overline{\mu}_c/J\to 1$, and for $\sigma/J\to \sqrt{2/\pi}$ one
finds $\overline{\mu}_c\to 0$. Thus for $\sigma\to 0$ we  recover
the ground state of the example
$P(\mu)=\delta[\mu-\overline{\mu}]$, as it should. For
$\overline{\mu}= 0$, on the other hand, we have
 $d \epsilon(p)/d p=p-{\rm Erf}[pJ/\sigma\sqrt{2}]$.
 Now one has only the solution $p=0$ (equal numbers of hydrophobic
 and hydrophilic monomers) for $\sigma/J> \sqrt{2/\pi}$,
and two equivalent stable solutions $\pm p$ with
$\sgn[p]=-\sgn[\overline{\mu}]$ for $\sigma/J< \sqrt{2/\pi}$,
separated by a second-order transition at $\sigma/J=
\sqrt{2/\pi}$.

 For $n\to\infty$ (deterministic genetic dynamics
one finds, upon doing the required integrals:
\begin{eqnarray}
 \lim_{n\to\infty} f_{\rm RS} & =&
 {\rm min}_{\{L_\phi\}}\left\{
 \frac{J}{4}\sum_{\phi}L^2_\phi
 -
 \frac{1}{2\beta}\log(\sum_\phi e^{\beta J
L_\phi}) -
 \frac{1}{2\beta}\log(\sum_\phi e^{-\beta J L_\phi}) \right.
\nonumber\\ &&\left. \hspace*{30mm}
 -F\left[\frac{1}{\sigma}\left(\overline{\mu}\minus
 \frac{1}{2\beta}\log\left[\frac{\sum_\phi e^{\beta
 JL_\phi}}{\sum_\phi e^{-\beta
 JL_\phi}}\right]\right)\right]
 \right\}
\end{eqnarray}
\begin{eqnarray}
p&=& -{\rm Erf}
\left[\frac{1}{\sigma\sqrt{2}}\left(\overline{\mu}\minus
\frac{1}{2\beta}\log\left[\frac{\sum_\phi e^{\beta
 JL_\phi}}{\sum_\phi e^{-\beta
 JL_\phi}}\right]\right)\right]
 \label{eq:gauss_ninfsaddle}
\end{eqnarray}
 with the function
$F[x]$ defined in (\ref{eq:defineF}), and together with our
order-parameter equation (\ref{eq:RSSPeqnsnew}) for the
$\{L_\phi\}$.  Here one will generally find a continuous
dependence of the average polarity $p$ on the system's control
parameters. Inspection of the swollen state $L_\phi=2p/q$
$\forall\phi$ here implies studying the saddle-point equation $p=
{\rm Erf}[(2pJ/q-\overline{\mu})/\sigma\sqrt{2}]$.
 For $\overline{\mu}=0$ one finds only the $p=0$ state for
 $\sigma/J>2\sqrt{2}/q\sqrt{\pi}$, which destabilizes at $\sigma/J=2\sqrt{2}/q\sqrt{\pi}$
 in favour of two energetically equivalent $p\neq 0$ solutions (with
 opposite signs).
For $\overline{\mu}\neq 0$ there is always a solution with
$\sgn[p]=-\sgn[\overline{\mu}]$, which always carries the lowest
free energy. A bifurcation analysis of our saddle-point equation
for $p$ reveals further that an alternative solution with
$\sgn[p]=\sgn[\overline{\mu}]$ is created discontinuously at
$\overline{\mu}=\pm \overline{\mu}_c$, where \bd
\frac{\overline{\mu}_c}{J}= \frac{2}{q}~{\rm
Erf}\left[\sqrt{\log\left(\frac{2\sqrt{2}J}{q\sigma\sqrt{\pi}}\right)}\right]
-\frac{\sigma\sqrt{2}}{J}\sqrt{\log\left(\frac{2\sqrt{2}J}{q\sigma\sqrt{\pi}}\right)}
\ed
 (provided $|\overline{\mu}|/J< (2/q){\rm Erf}[\sqrt{2/\pi}]$).

As with binary distributed $\mu$, again we find that for
sufficiently large $\sigma$ the single-species states $p=\pm 1$
are replaced by biologically more interesting states with equal
fractions of hydrophilic  and hydrophobic monomers.

\subsection{Phase diagrams for $q=2$}

\begin{figure}[t]
\vspace*{23mm}\hspace*{1mm} \setlength{\unitlength}{0.62mm}
\begin{picture}(0,70)
\put(0,-10){\epsfxsize=140\unitlength\epsfbox{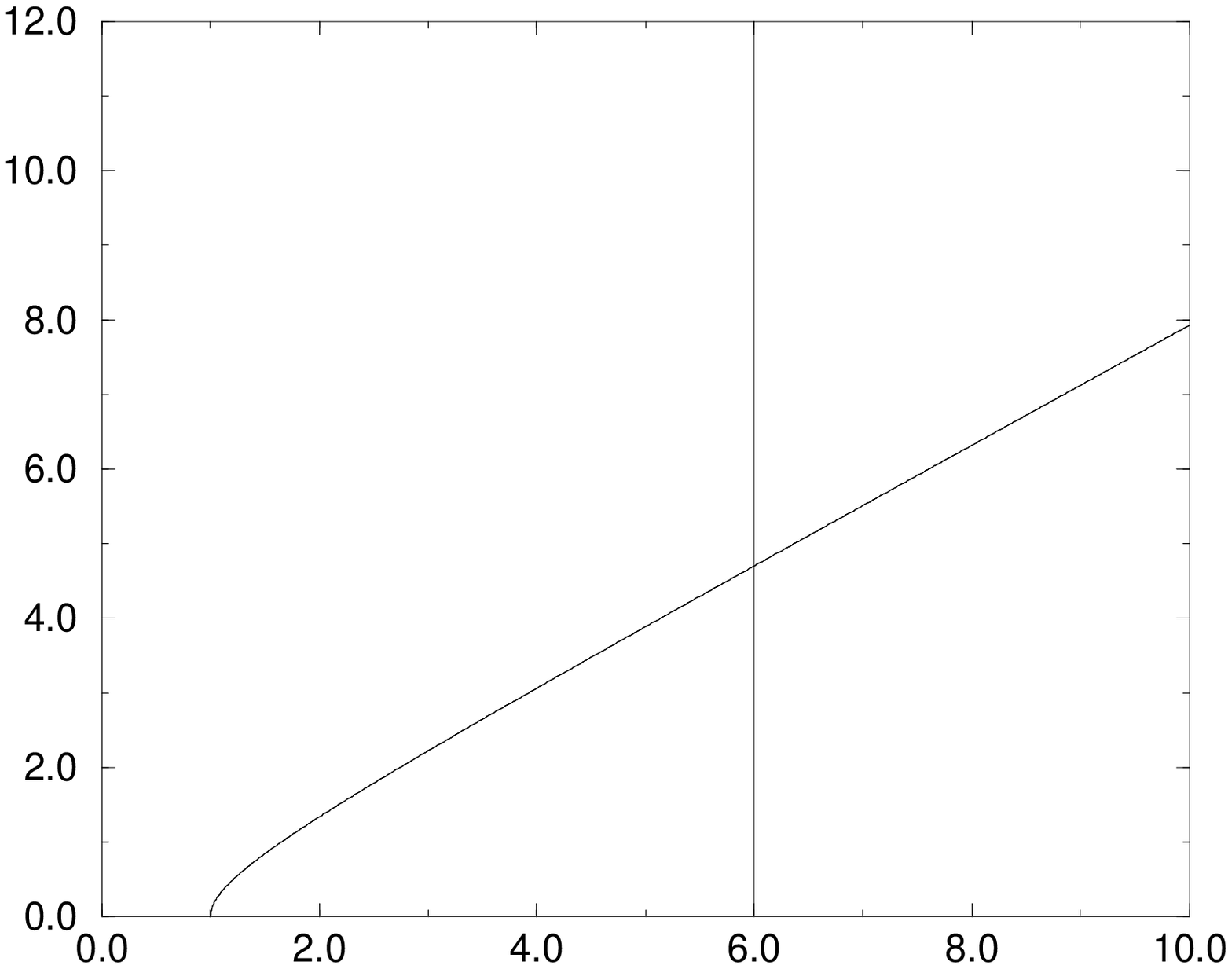}}
\put(0, 50){\large $\tilde{\beta}\sigma$} \put(69,-9){\large
$\tilde{\beta} J$} \put( 50,  60){$S1$}  \put(68, 20){$S2$}\put(
100, 78){$C1$}  \put(103, 30){$C2$}
\put(140,-10){\epsfxsize=140\unitlength\epsfbox{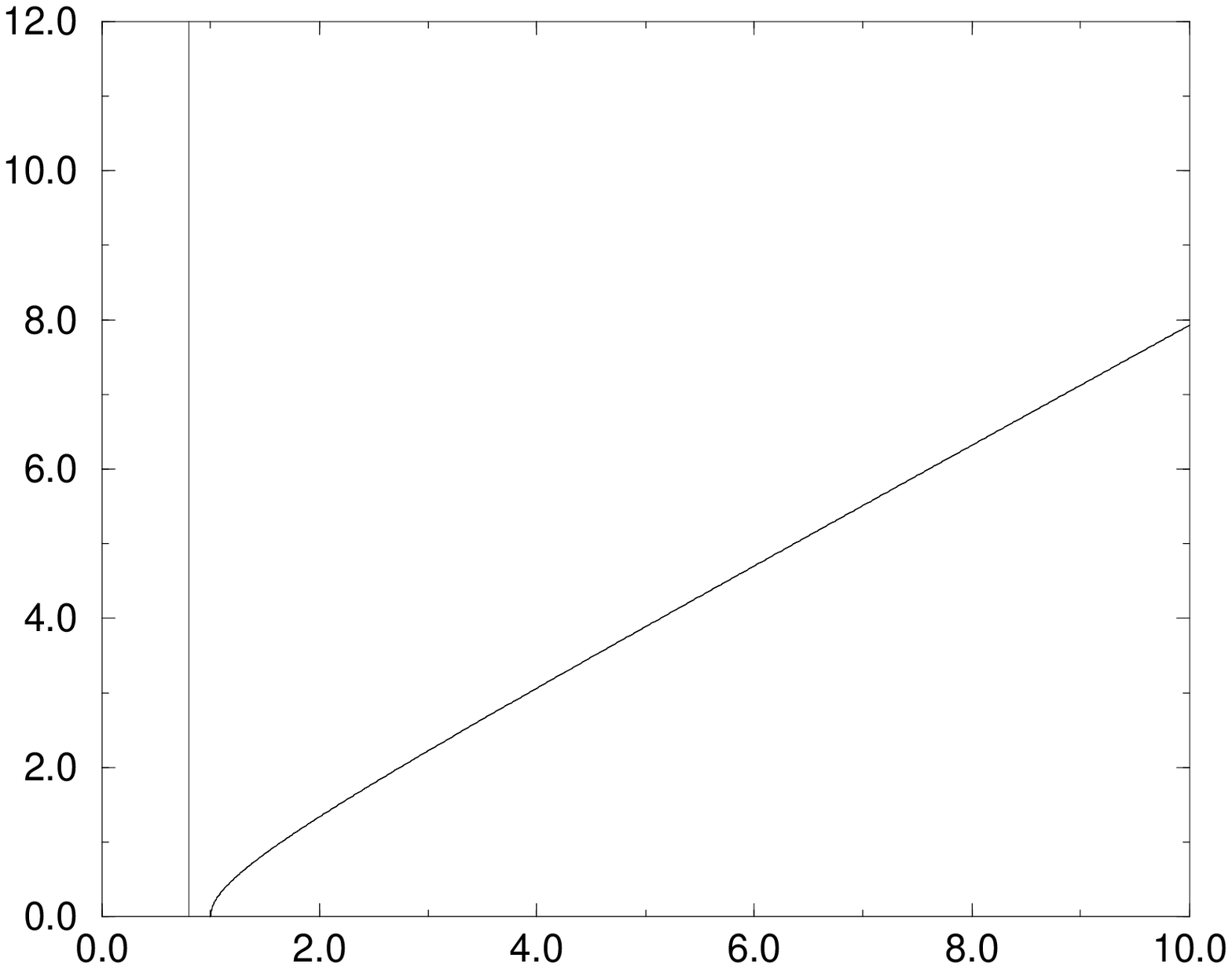}}
\put(140,  50){\large $\tilde{\beta}\sigma$} \put(209,-9){\large
$\tilde{\beta} J$} \put(164,50){$S1$}\put( 200, 60){$C1$}
\put(233, 25){$C2$}
\end{picture}
\vspace*{10mm} \caption{Phase diagrams for $q=2$ and
$P(\mu)=[2\pi\sigma^2]^{-1/2}e^{-\frac{1}{2}\mu^2/\sigma^2}$. The
four possible phases are $\{S1,S2\}$ (swollen phases with $1$ or
$2$ possible locally stable values of $p$) and $\{C1,C2\}$
(compact phases with $1$ or $2$ possible locally stable values of
$p$). Here all transitions are second order.  Left diagram: $n=6$.
 Right diagram: $n=\frac{3}{4}$. Note that $\tilde{\beta}=\beta
n$, and that
 phase $S2$ exists only for $n>1$.} \label{fig:gaussianphasediagramq=2}
\end{figure}

Working out the $q=2$ equations (\ref{eq:saddle_q2}) for
zero-average Gaussian genetic forces (since choosing
$\overline{\mu}\neq 0$ always has a deteriorating effect) leads us
to the following saddle-point equation for the polarity order
parameter $p$ (with the short-hand
$Dz=(2\pi)^{-\frac{1}{2}}e^{-\frac{1}{2}z^2}$): \be
p=G(p),~~~~~~~~ G(p)=\int\!Dz~ \tanh[\tilde{\beta}(Jp\minus \sigma
z)] \label{eq:peqn_gaussian} \ee  Again  $p=0$ is always a
solution. We expand $G(p)$ for small $p$ and find \bd
G(p)=\tilde{\beta}J p\int\!\!Dz\left[1\minus
\tanh^2(\tilde{\beta}\sigma z)\right]\ed \bd\hspace*{20mm} -
\frac{1}{3}(\tilde{\beta}J)^3p^3\int\!\!Dz\left[ 1\minus
4\tanh^2(\tilde{\beta}\sigma z)\plus 3\tanh^4(\tilde{\beta}\sigma
z)\right]+\order(p^5) \ed and conclude that the state $p=0$
de-stabilizes at \be
1=\tilde{\beta}J\int\!Dz\left[1-\tanh^2(\tilde{\beta}\sigma z
)\right]
 \label{eq:second_order_gauss} \ee
We note that \begin{eqnarray*}
\tilde{\beta}\sigma_c=(\tilde{\beta}J\minus 1)^{1/2}
+\order(\tilde{\beta}J\minus 1) &~~~~~~~& (\tilde{\beta}J\to 1)\\
\tilde{\beta}\sigma_c=\sqrt{\frac{2}{\pi}}\tilde{\beta}
J+\order(1) & ~~~~~~~& (\tilde{\beta}J\to \infty)
\end{eqnarray*}
(in line with our earlier results on the ground state).
 The transition (\ref{eq:second_order_gauss})
would be preceded by a first order one from the point onwards
where $G^{\prime\prime\prime}(0)\geq 0$. Along the line
(\ref{eq:second_order_gauss}) this latter condition translates
into \bd 1\leq \frac{3}{4}\tilde{\beta}J\int\!\!Dz\left[ 1\minus
\tanh^4(\tilde{\beta}\sigma z)\right]
  \ed
Numerical analysis reveals that this inequality is never
satisfied, and there is no evidence for first order transitions.
The final picture for $q=2$ is thus as follows.
 There are only second order transitions: one  occurs
 at $\beta J=1$ from a state
with uniformly distributed monomer orientations (representing a
swollen state) to a state with separation of polarity types
(representing a compact state), the remaining two occur for
$\sigma=\pm \sigma_c$, where $\sigma_c$ denotes the non-negative
solution of
 (\ref{eq:second_order_gauss}),
and mark the creation of two locally stable non-zero values for
the average polarity (which differ in sign only) together with
de-stabilization of the state $p=0$. These lines are shown in the
$(\tilde{\beta}J,\tilde{\beta}\sigma)$ phase diagram, in figure
\ref{fig:gaussianphasediagramq=2}. We again denote swollen phases
with $\ell$ locally stable values of $p~$ as $S\ell$, and
 compact phases with $\ell$  locally stable values of
$p~$ as $C\ell$. As in the previous examples, phase $S2$ exists
only for $n>1$.

\subsection{Phase diagrams for $q=3$}

\begin{figure}[t]
\vspace*{28mm}\hspace*{41mm} \setlength{\unitlength}{0.62mm}
\begin{picture}(0,70)
\put(0,-10){\epsfxsize=140\unitlength\epsfbox{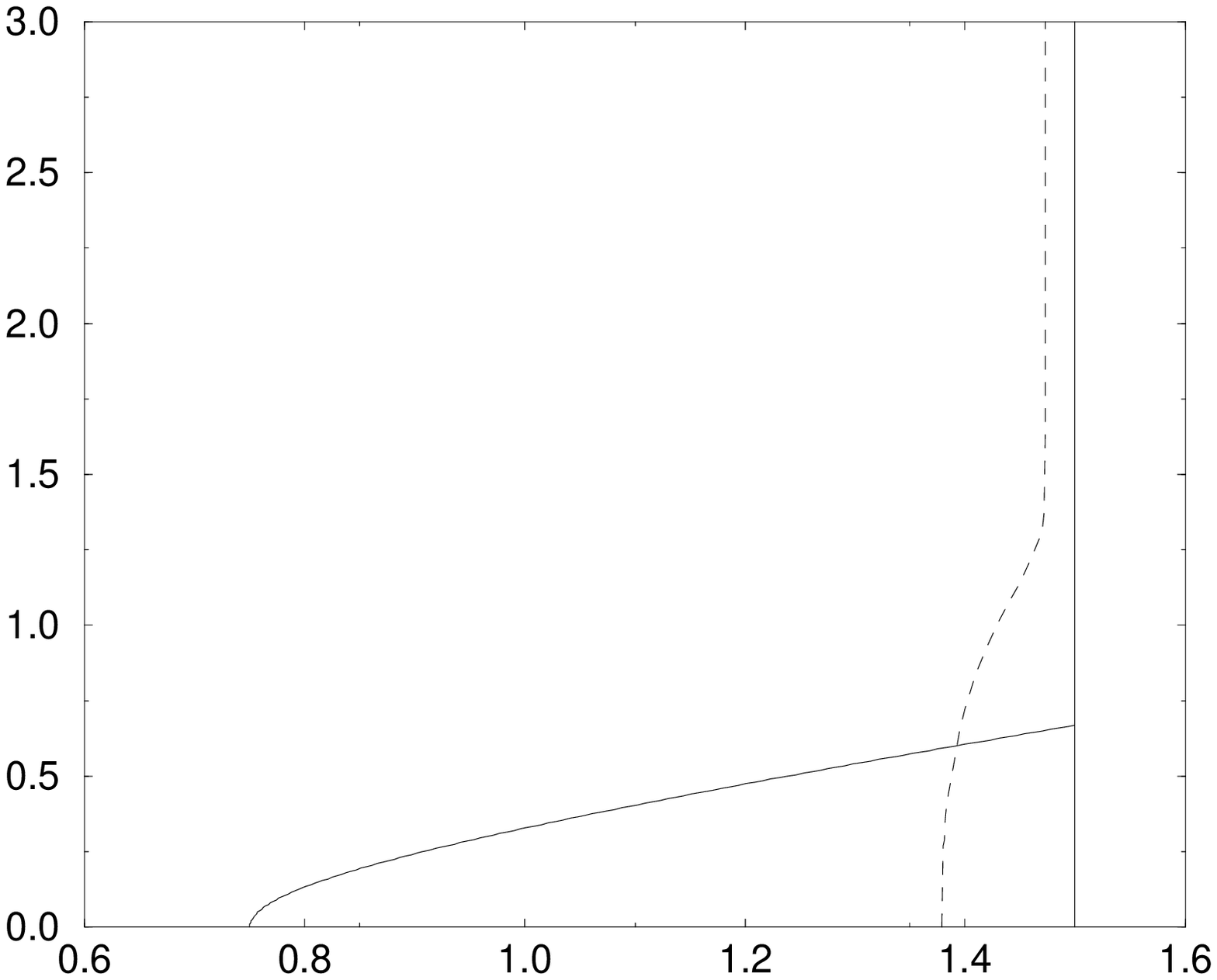}}
\put( 0, 48){\large $\tilde{\beta}\sigma$} \put( 70,  -9){\large
$\beta{J}$} \put( 40,  70){$S1$}  \put( 80, 13){$S2$} \put( 117,
40){$CO$}
 \put(80,60){$S1+C1$}
 \put(104,60){\line(1,-1){10}}
  \put(75,32){$S2+C2$}
 \put(99,32){\line(1,-1){10}}
\end{picture}
\vspace*{8mm} \caption{ Phase diagram for $q=3$ and
$P(\mu)=[2\pi\sigma^2]^{-\frac{1}{2}}e^{-\frac{1}{2}\mu^2/\sigma^2}$,
with $n=2$. The possible phases are $\{S\ell\}$ (swollen phases
with $\ell$ possible locally stable values of $p$), $\{C\ell\}$
(compact phases with $\ell$ possible locally stable values of
$p$), $\{S\ell\plus C\ell\}$ (phases with $2\ell$ possible locally
stable states of $p$, half of which are compact), and $CO$
(compact states only). First order transitions are indicated by
dashed curves, second order ones by solid curves. }
\label{fig:gaussianq3}
\end{figure}

For the force distribution
$P(\mu)=(2\pi\sigma^2)^{-\frac{1}{2}}e^{-\frac{1}{2}\mu^2/\sigma^2}$
 we
know the ground state to be $\bL=(-1,0,1)$ for
$\sigma>J\sqrt{2/\pi}$ (with $p=0$) and $\bL=2p(1,0,0)$ for
$\sigma<J\sqrt{2/\pi}$ (with $p=\pm 1$). We therefore expect the
solution to be of the relatively form $\bL=(L_1,L_2,L_2)$ at all
temperatures {\em at most} when $\sigma<J/2$. Inserting $\{L_{\pm
2\pi/3}=\frac{1}{3}(2p+Z),L_0=\frac{1}{3}(2p-2Z)\}$ into our
saddle-point equations now leads to
\be
p= \int\!Du \left[\frac{e^{\beta n[\sigma u
+\frac{2}{3}Jp]}(2e^{\frac{1}{3}\beta J Z}+e^{-\frac{2}{3}\beta J
Z})^n- e^{-\beta n[\sigma u+\frac{2}{3}Jp]}(2e^{-\frac{1}{3}\beta
J Z}+e^{\frac{2}{3}\beta J Z})^n }{e^{\beta n[\sigma u
+\frac{2}{3}Jp]}(2e^{\frac{1}{3}\beta J Z}+e^{-\frac{2}{3}\beta J
Z})^{n}+ e^{-\beta n[\sigma u+\frac{2}{3}Jp] }
(2e^{-\frac{1}{3}\beta J Z}+e^{\frac{2}{3}\beta J Z})^n} \right]
\ee
\be
Z=F[Z;p]~~~~~~~~~~F[Z;p]=\frac{2p[1-\cosh(\beta JZ)]+6\sinh(\beta
JZ)}{5+4\cosh(\beta JZ)} \ee
 For $Z=0$ these equations bring us
back to the swollen state $L_\phi=\frac{2}{3}p$, but now with \bd
p=\int\!Dz~\tanh[\beta n(\frac{2}{3}Jp+\sigma z)]
 \ed Comparison with the
corresponding equation for $q=2$ shows that, for
$\sigma<J\sqrt{2/\pi}$, the properties of the
 swollen state with $q=3$ can again be obtained from those derived for
$q=2$ via the substitution $J\to \frac{2}{3}J$. This tells us that
the state $p=0$ (which always solves the saddle-point equation)
de-stabilizes at
\be
1=\frac{2}{3}\tilde{\beta}J
\int\!Dz\left[1-\tanh^2(\tilde{\beta}\sigma z)\right]
\label{eq:2nd_order_gauss_q3} \ee and that this corresponds to a
true second-order transition.

We show the resulting phase diagram (complemented by numerical
analysis of the transitions) in figure \ref{fig:gaussianq3}, for
$n=2$. As a result of the symmetry $P(\mu)=P(\minus \mu)$ the $S1$
phase always has $p=0$ (in contrast to e.g.  $\delta$-distributed
forces).
 The $S2$ and $C2$ phases involve $p\neq 0$ (with the two  values different in sign only).
 One notes again the crucial dependence on $n$ of the
existence of some of the phases (clearly evident in the equations
above). It is  clear that upon choosing a sufficiently large value
for $\sigma$ one can eliminate the biologically undesirable single
species states at any temperature $T$.

\section{Discussion}

Our objective  was to define and study a simple solvable model
describing the coupled dynamics of (fast) compactification and
(slow) genetic monomer sequence selection in hetero-polymers,
driven by the competing demands of functionality and
reproducibility of the resulting folded structures, as a first
step towards understanding the genesis and statistics of natural
amino-acid sequences in proteins. Our model is a simple mean-field
hetero-polymer whose state is described by two degrees of freedom
per monomer $i$: an angle $\phi_i$, giving its orientation
relative to the backbone, and a binary variable $\eta_i$, giving
its polarity (i.e. hydrophobic vs. hydrophilic). The evolution of
the orientation variables represents folding; that of the polarity
variables (which involves changing monomer species) represents
genetic evolution. The latter is assumed to be adiabatically slow
compared to the former. There is also explicit (quenched) site
disorder in the model, in the form of a simple (site-factorized)
random functionality potential for the monomer sequences.

We have solved our model in equilibrium using the finite-$n$
replica formalism, designed to analyze statistical mechanical
systems with disparate time-scales, where $n$ denotes the ratio of
the noise levels (or temperatures) in the two stochastic
processes. Replica symmetry is stable, and we are able to derive
closed equations for the system's order parameters, which measure
the degree of separation of the monomer species, and can be
interpreted in terms of swollen versus compact states. This leads
to explicit analytical results for ground states and
high-temperature states. Solution of our order-parameter equations
at intermediate temperatures requires a combination of analytical
and  numerical methods. Since the qualitative behaviour of the
system can already be captured by restricting ourselves to $q\leq
3$ ($q$ denotes the number of possible orientation angles per
monomer), we calculate phase diagrams for $q\in\{2,3\}$ and for
different choices of the disorder in the sequence functionality
potential. These exhibit a rich phenomenology of first- and
second-order phase transitions, and shed light on the emergence of
monomer species statistics (characterized by the fraction of
hydrophobic versus hydrophilic monomers) from the underlying
genetic dynamics. The properties of the functionality potential
are found to have a large impact on the nature of the phase
diagram; the distribution of genetic forces needs to be
sufficiently broad relative to its average, in order to prevent
the system from evolving towards a single-species state (where all
monomers have the same polarity). This is perfectly compatible
with the efficiency-driven biological need for large sequence
diversity (in order to use the available hetero-polymer hardware
for as many different biological functions as possible), which
again favors the $p=0$ states.

There is much scope for further work. At a theoretical level one
could repeat the above analysis for models in which the folding
process also includes short-range forces (e.g. hydrogen bonds and
steric forces), as in \cite{Skantzos}), which would involve
$n$-replicated transfer matrices, or study the dynamics (at either
the fast or the slow time-scale, or both). One could also inspect
more realistic sequence functionality potentials $V(\bfeta)$ with
a large number of local minima. At the level of increasing
biological realism one could introduce more sophisticated degrees
of freedom for the monomers, such as three real-valued orientation
angles per monomer, or allow for a realistic number of monomer
species (as opposed to two) and endow each with additional
physical characteristics.

A final criticism to be raised against the present model is that
its genetic Hamiltonian favours sequences which allow for
efficient shielding of hydrophobic monomers and which meet
functionality requirements. Efficient shielding, and hence
reliable compactification, is our simplified measure of structure
reproducibility. It would be interesting to construct and solve a
model in which the genetic Hamiltonian really measures the number
of meta-stable states, and rewards sequences which are not only
guaranteed to generate a compact conformation (as in the present
model) but also a unique one.

\clearpage
\section*{References}

\appendix

\section{Stability analysis}
\label{app:stability}

\subsection*{General properties}

Here we analyse the local stability properties of the
saddle-points of the free energy surface per monomer $f[\{z\}]$,
which is extremized in (\ref{eq:replicatedfenergy}), by studying
the eigenvalue problem of the Hessian
$D_{\alpha\phi,\gamma\psi}=\partial^2 f[\{z\}]/\partial
z_\phi^\alpha
\partial z_{\psi}^\gamma$. In replica-symmetric saddle-points, where
$z_\phi^\alpha=z_\phi=\beta J L_\phi$ for all $\alpha$, the
Hessian  takes the following form
\be
D_{\alpha\phi,\gamma\psi}=\frac{1}{\tilde{\beta}}\left\{
\frac{1}{2{\beta}J}\delta_{\alpha\gamma}
\delta_{\phi\psi}+\tilde{D}_{\alpha\phi,\gamma\psi}\right\}
\label{eq:hessiansplit} \ee
 \be
\tilde{D}_{\alpha\phi,\gamma\psi}=
\hat{K}_{\phi\psi}-\hat{L}_{\phi\psi}+\delta_{\alpha\gamma}[
\hat{L}_{\phi\psi}-\hat{M}_{\phi\psi}]
 \label{eq:operatorsplit} \ee
where
\begin{eqnarray}
\hat{K}_{\phi\psi}&=& \bra
 [w_+ v^+_\psi \minus w_- v^{-}_\psi ]
 [w_+ v^+_\phi \minus w_- v^{-}_\phi ]
 \ket_\mu
 \label{eq:hatK}
\\
 \hat{L}_{\phi\psi}&=&
 \bra w_+ v^+_\psi v^+_\phi + w_- v^-_\psi v^-_\phi \ket_\mu
 \label{eq:hatL}
\\
\hat{M}_{\phi\psi}&=& \delta_{\phi\psi}\bra
 w_+ v^+_\psi
 + w_- v^-_\psi \ket_\mu
 \label{eq:hatM}
\end{eqnarray}
and with the short-hands
 \bd
w_{\pm}=\frac{e^{\mp\tilde{\beta}\mu}(\sum_\phi e^{\pm \beta J
L_\phi})^n} {e^{-\tilde{\beta}\mu}(\sum_\phi e^{\beta J
L_\phi})^n+ e^{\tilde{\beta}\mu}(\sum_\phi e^{- \beta J
L_\phi})^n}~~~~~~~~~~~~~ v^\pm_\phi=\frac{e^{\pm \beta J
L_{\phi}}}
 {\sum_{\phi^\prime} e^{\pm \beta J L_{\phi^\prime}}}
\ed The three matrices $\{\hat{K},\hat{L},\hat{M}\}$ are all
positive definite and symmetric, and the non-negative quantities
$w_\pm$ and $v^\pm_\phi$ obey the  normalization relations $w_+
+w_-=1$ and $\sum_\phi v^\pm_\phi=1$. For high temperatures (where
we know the system is ergodic) we just find \bd \beta^2
D_{\alpha\phi,\gamma\psi}=\frac{1}{n}\left\{
\frac{1}{2J}\delta_{\alpha\gamma}
\delta_{\phi\psi}+\order(\beta)\right\}~~~~~~~~ (\beta\to 0 )
 \ed (i.e. $D$ is positive
definite for $\beta\to 0$). Hence the condition for a local
de-stabilization of an RS saddle-point $\{L_\phi\}$ is given in
terms of the smallest eigenvalue of $\tilde{D}$ as
\be
\sum_{\gamma\psi} \tilde{D}_{\alpha\phi,\gamma\psi}
x_\psi^\gamma=\lambda x_\phi^\alpha, ~~~~~~~~~~ \lambda_{\rm
min}=-1/2 \beta J \label{eq:second_order_trans}
 \ee
We now set out to determine the eigenvectors and eigenvalues of
(\ref{eq:operatorsplit}).

\subsection*{RS \& RSB eigenvectors, local stability of the RS saddle-point}

Replica-symmetric eigenvectors, describing fluctuations within the
RS subspace, are of the form  $x^{\alpha}_\phi=x_{\phi}$ for all
$\alpha$. Insertion into the eigenvalue problem of
(\ref{eq:operatorsplit}) gives
\be
\left[n(\hat{\bK}-\hat{\bL})+\hat{\bL}-\hat{\bM}\right]\bx=\lambda_{\rm
RS}~\bx \label{eq:RSeigenvalueidentity} \ee with $\bx=\{x_\phi\}$.
Since the matrices (\ref{eq:hatK},\ref{eq:hatL},\ref{eq:hatM}) are
$q\times q$ ones, we know that the RS eigenspace is
$q$-dimensional. The matrix $\tilde{D}$ is self-adjoint, so we
know that the RSB eigenvectors, which we write as
$y^{\alpha}_\phi$, must be orthogonal to the above $q$-dimensional
RS eigenspace; they must obey the $q$ conditions
$\sum_{\alpha}y^{\alpha}_\phi=0$ for all $\phi$. Insertion into
the eigenvalue problem of (\ref{eq:operatorsplit}) now gives
\be
\sum_{\psi}[ \hat{L}_{\phi\psi}-\hat{M}_{\phi\psi}] y_\psi^\alpha
=\lambda_{\rm RSB}~y_\phi^\alpha
 \ee
Together with the othogonality conditions this leaves an $(n-1)q$
dimensional RSB eigenspace, as it should. We conclude that, in
view of (\ref{eq:second_order_trans}), that second-order
transitions of the RS and RSB type occur when, respectively
\be
{\rm RS~transitions:}~~~~~~~~\min_{\bx\in \real^q,~\neq
\bnull}\left[ \frac{\bx\cdot[n(\hat{\bK}\minus
\hat{\bL})+\hat{\bL}-\hat{\bM}]\bx}{\bx^2}\right] =-\frac{1}{2
\beta J} \ee
\be
{\rm RSB~transitions:}~~~~~~~~\min_{\bx\in \real^q,~\neq \bnull}
\left[\frac{\bx\cdot[\hat{\bL}-\hat{\bM}]\bx}{\bx^2}\right]
=-\frac{1}{2 \beta J} \ee Below we show that the matrix
$\hat{K}\minus \hat{L}$ is negative definite. As a consequence we
can be certain that a de-stabilization of an RS saddle-point will
always occur within the RS eigenspace, since \bd \min_{\bx\in
\real^q,~\neq\bnull}\left[ \frac{\bx\cdot[n(\hat{\bK}\minus
\hat{\bL})+\hat{\bL}-\hat{\bM}]\bx}{\bx^2}\right]< \min_{\bx\in
\real^q,~\neq\bnull}\left[
\frac{\bx\cdot[\hat{\bL}-\hat{\bM}]\bx}{\bx^2}\right] \ed (only
non-physical RS saddle-points can be locally unstable against RSB
fluctuations). Hence replica-symmetry will always be locally
stable. What remains is to confirm that $\hat{K}\minus \hat{L}$ is
indeed negative definite. To do so we first define $ \bra
x\ket_\pm=\sum_\phi v^{\pm}_\phi x_\phi$. This allows us to write
 \begin{eqnarray*}
\sum_{\phi\phi^{\prime}}x_{\phi}[\hat{K}_{\phi \phi^{\prime}}-
\hat{L}_{\phi\phi^{\prime}}] x_{\phi^{\prime}}&=&
 \bigbra~
\left[ w_+ \bra x\ket_+ -w_- \bra x \ket_- \right]^2
-
 w_+ \bra x\ket_+^2 - w_- \bra x\ket_-^2
 ~\bigket_{\!\!\mu}
\\
& \leq&
 \bigbra~
\left[ w_+ |\bra x\ket_+| +w_- |\bra x \ket_-| \right]^2
-
 w_+ |\bra x\ket_+|^2 - w_- |\bra x\ket_-|^2
 ~\bigket_{\!\!\mu}\\
  &\leq & 0
\end{eqnarray*}
This completes our proof.


\begin{thebibliography}{99}
\bibitem{PDB}
Research Collaboratory for Structural Bioinformatics (NSF, DoE,
NIH) 1987 {\em Protein Data Bank}, {\tt
http://www.pdb.bnl.gov/index.html}

\bibitem{PIR}
National Biomedical Research Foundation (USA), Munich Information
Center for Protein Sequences, Japanese International Protein
Sequence Database 1984 {\em Protein Information Resource}, {\tt
http://www-nbrf.georgetown.edu/pirwww/pirhome.shtml}

\bibitem{MDreview}
Brooks C 1998 {\em Curr.~Opinion.~Struct.~Biol.} {\bf 8}   222

\bibitem{Dill}
Dill K A, Bromberg S, Yue K, Fiebig K M, Yee D P, Thomas P D and
Chan H S 1995 {\em Protein Science} {\bf 4} 561

\bibitem{Pande}
Pande V and Roksar D 1999 {\em Proc.~Natl.~Acad.~Sci.~USA} {\bf
96} 9062

\bibitem{Wang}
Wang T, Miller J, Wingreen N S, Tang C and Dill K A 2000 {\em
J.~Chem.~Phys.} {\bf 113} 8329

\bibitem{GOT}
Garel T, Orland H and Thirumalai D 1996 in {\em New Developments
in Theoretical Studies of Proteins} R Elber Ed. (Singapore: World
Scientific)

\bibitem{GOP}
Garel T, Orland H and Pitard E 1998 in {\em Spin Glasses and
Random Fields} A P Young Ed. (Singapore: World Scientific)

\bibitem{CPS}
Coolen A C C, Penney R W and Sherrington D 1993
   {\it Phys.~Rev.~B} {\bf 48} 16116

\bibitem{PCS}
Penney R W, Coolen A C C and Sherrington D 1993
   {\it J.~Phys.~A: Math.~Gen.} {\bf 26} 3681

\bibitem{DFM}
Dotsenko V, Franz S and M\'ezard M 1994
   {\it J.~Phys.~A: Math.~Gen.} {\bf 27} 2351

\bibitem{PS}
Penney R W and Sherrington D 1994
   {\it J.~Phys.~A: Math.~Gen.} {\bf 27} 4027

\bibitem{Caticha}
Caticha N 1994
   {\it J.~Phys.~A: Math.~Gen.} {\bf 27} 5501

\bibitem{Jongenetal2}
Jongen G, Anem\"{u}ller J, Boll\'{e} D, Coolen A C C and
P\'{e}rez-Vicente C J 2000
   {\it J. Phys.~A: Math. Gen.} {\bf 34} 3957

\bibitem{MourikCoolen}
Van~Mourik J and Coolen A C C 2001
   {\it J.~Phys.~A: Math.~Gen.} {\bf 34} L111

\bibitem{UezuCoolen}
Uezu T and Coolen A C C 2002  {\it J.~Phys.~A: Math.~Gen.} {\bf
35} 2761

\bibitem{Skantzos}
Skantzos N S, van Mourik J and Coolen A C C 2000 {\em J.~Phys.~A:
Math.~Gen.} {\bf 34} 4437

\bibitem{Garel}
Garel T and  Orland H 1988
{\em Europhys.~Lett.} {\bf 6} 597

\bibitem{SG}
Shakhnovich E I and Gutin A M 1989 {\em J.~Phys.~A: Math.~Gen.}
{\bf 22} 1647

\bibitem{Stafos}
 Stafos C D, Gutin A M and  Shakhnovich E I 1993
{\em Phys.~Rev.~E} {\bf 48} 465

\bibitem{Mattis}
Mattis D C 1976 {\em Phys.~Lett.} {\bf 56A} 421

\end{thebibliography}
\end{document}